\documentclass[journal]{IEEEtran}
\usepackage[latin9]{inputenc}
\usepackage{geometry}
\usepackage{color}
\usepackage{float}
\usepackage{amsmath}
\usepackage{amssymb}
\usepackage{graphicx}

\makeatletter

\providecommand{\tabularnewline}{\\}
\floatstyle{ruled}
\newfloat{algorithm}{tbp}{loa}
\providecommand{\algorithmname}{Algorithm}
\floatname{algorithm}{\protect\algorithmname}

\usepackage{amsfonts}
\usepackage{mathrsfs}
\usepackage{mathrsfs}\usepackage[noblocks]{authblk}
\usepackage[compress]{cite}
\usepackage{bm}
\usepackage{algorithm}
\usepackage{algorithmic}
\usepackage{epsfig}\usepackage{graphics}\usepackage{subfigure}\usepackage{epsfig}\usepackage{epstopdf}
\usepackage{graphics}\usepackage{subfigure}\usepackage{upgreek}\usepackage{caption}\usepackage{theorem}
\usepackage{breqn}
\usepackage{multicol}
\usepackage{eufrak}
\usepackage{eucal}
\usepackage{array}
\usepackage[compress]{cite}
\usepackage{algorithm}
\usepackage{algorithmic}

\allowdisplaybreaks[4]

\newtheorem{theorem}{Theorem}\newtheorem{lemma}{Lemma}\theoremheaderfont{\normalfont\bfseries}

\makeatother

\begin{document}
\title{A Framework of Robust Transmission Design for IRS-aided MISO Communications
with Imperfect Cascaded Channels}
\author{Gui~Zhou, Cunhua~Pan, Hong~Ren, Kezhi~Wang, and Arumugam~Nallanathan,~\IEEEmembership{Fellow, IEEE}
 \thanks{(Corresponding author: Cunhua Pan) G. Zhou, C. Pan, H. Ren and A.
Nallanathan are with the School of Electronic Engineering and Computer
Science at Queen Mary University of London, London E1 4NS, U.K. (e-mail:
g.zhou, c.pan, h.ren, a.nallanathan@qmul.ac.uk). K. Wang is with Department
of Computer and Information Sciences, Northumbria University, UK.
(e-mail: kezhi.wang@northumbria.ac.uk). This work was supported by
Grant EP/R006466/1. }}
\maketitle
\begin{abstract}
Intelligent reflection surface (IRS) has recently been recognized
as a promising technique to enhance the performance of wireless systems
due to its ability of reconfiguring the signal propagation environment.
However, the perfect channel state information (CSI) is challenging
to obtain at the base station (BS) due to the lack of radio frequency
(RF) chains at the IRS. Since most of the existing channel estimation
methods were developed to acquire the cascaded BS-IRS-user channels,
this paper is the first work to study the robust beamforming based
on the imperfect cascaded BS-IRS-user channels at the transmitter
(CBIUT). Specifically, the transmit power minimization problems are
formulated subject to the worst-case rate constraints under the bounded
CSI error model and the rate outage probability constraints under
the statistical CSI error model, respectively. After approximating
the worst-case rate constraints by using the S-procedure and the rate
outage probability constraints by using the Bernstein-type inequality,
the reformulated problems can be efficiently solved. Numerical results
show that the negative impact of the CBIUT error on the system performance
is greater than that of the direct CSI error. 
\end{abstract}


\section{Introduction}

Intelligent reflecting surface (IRS), which is also known as reconfigurable
intelligent surface (RIS) or large intelligent surface (LIS), has
emerged as a promising technique to enhance the spectral and energy
efficiency of the wireless networks \cite{Marco-4,Marco-3,Xiaojun},
thanks to its artificial planar passive radio array structure which
is cost-effective and energy-efficient. More explicitly, each passive
element on the IRS is capable of reconfiguring the channels between
the BS and users constructively or destructively by imposing an independent
phase shift to the incident signal. The existing literature on IRS-aided
wireless communications has demonstrated that IRS is an enabler for
enhancing the spectral and energy efficiency through jointly optimizing
the active beamforming at the BS and the passive beamforming at the
IRS \cite{Pan2019intelleget,Pan2019multicell,Baitong2019,Marco-2,Gui2019IRS,Xianghao2009,Shen2019secrecy,OFDM2019}.
However, the algorithms developed in the above contributions were
based on the assumption of perfect channel state information at the
transmitter (CSIT).

Unfortunately, it is challenging to estimate the channels for the
IRS-aided wireless systems, since IRS is passive and can neither send
nor receive pilot symbols. In IRS-aided communication systems, there
are two types of channels: the direct channel spanning from the BS
to the user, and the IRS-related channels. The direct channel can
be readily estimated by using conventional channel estimation methods
such as the least square algorithm. Hence, most of the existing contributions
focused on the channel estimation for the IRS-related channels, which
are composed of the channel from the BS to the IRS (BS-IRS channel),
and those from the IRS to the users (IRS-user channels).

In general, there are two main approaches to estimate the IRS-related
channels. The first approach is to directly estimate the IRS-related
channels, i.e., estimate BS-IRS channel and IRS-user channels separately
\cite{taha2019enabling}. Specifically, in \cite{taha2019enabling},
some active channel elements are installed at the IRS to estimate
the individual channels. This method, however, has several drawbacks.
The active elements may increase the hardware cost and consume extra
power, which causes unaffordable burden on the IRS. In addition, the
channel information estimated at the IRS needs to be fed back to the
BS, which increases the information exchange overhead.

Fortunately, it is observed that the cascaded BS-IRS-user channels,
which are the product of the BS-IRS channel and the IRS-user channels,
are sufficient for the joint active and passive beamforming design
\cite{Gui2019IRS,Xianghao2009,Shen2019secrecy,OFDM2019}. As a result,
most of the existing contributions focused on the second approach,
i.e., the cascaded channel estimation \cite{zhengyi-est,shuguang-IRS,Peilan,chenjie}.
Specifically, the channel estimation of the cascaded channel has been
investigated both in the single-user multiple-input multiple-output
(SU-MIMO) system \cite{zhengyi-est} and the multi-user multiple-input
single-output (MU-MISO) system \cite{shuguang-IRS}. However, the
pilot overhead of the estimation methods in \cite{zhengyi-est,shuguang-IRS}
is prohibitively high, which scales up with the number of reflection
elements. In order to reduce the pilot overhead, the authors in \cite{Peilan}
exploited the sparse property of the channel matrix and proposed a
channel estimation method based on compressed sensing technique. Furthermore,
another sparsity representation of the cascaded channel has been found
in \cite{chenjie} by using the fact that the height of the BS and
the IRS are often the same.

All the above-mentioned literature \cite{Pan2019intelleget,Pan2019multicell,Baitong2019,Marco-2,Gui2019IRS,Xianghao2009,Shen2019secrecy,OFDM2019}
did not consider the transmission design by taking into account the
channel estimation error. Due to the inevitable channel estimation
error, it will induce system performance loss if naively treating
the estimated channels as perfect ones. Hence, it is imperative to
design robust transmission strategies for the IRS-aided wireless communication
systems. To the best of our knowledge, there are only a few contributions
in this area \cite{Gui-letter,xianghao-robust}. Specifically, in
\cite{Gui-letter}, we first proposed a worst-case robust design algorithm
by assuming that the BS only knew the imperfect IRS-user channels
in a MU-MISO wireless system. Then, the authors in \cite{xianghao-robust}
further proposed a robust secure transmission strategy by also applying
the worst-case optimization method when the channels from the IRS
to the eavesdroppers were imperfect. However, to implement the above
robust design algorithms in \cite{Gui-letter} and \cite{xianghao-robust},
one should rely on the first channel estimation approach, where the
BS-IRS channels and IRS-user channels should be independently estimated.
This is difficult to achieve since several active elements should
be installed at the IRS.

Against the above background, this paper studies the robust transmission
design based on the imperfect cascaded BS-IRS-user channels at the
transmitter (CBIUT)\footnote{Part of this work is published in \cite{Gui-globecom}.}.
Specifically, we aim to design a robust active and passive beamforming
scheme to minimize the total transmit power under both the bounded
CSI error model and the statistical CSI error model. Unfortunately,
the robust beamforming algorithms developed in \cite{Gui-letter}
and \cite{xianghao-robust} are not applicable for the imperfect CBIUT
case. Hence, the contributions of this work are summarized as follows: 
\begin{itemize}
\item To the best of our knowledge, this is the first work to study the
robust transmission design based on imperfect cascaded BS-IRS-user
channels, which is more practical than the previous works in which
imperfect IRS-user channels were considered. In addition, we consider
the robust transmission design under two channel error models: the
bounded CSI error model and the statistical CSI error model. However,
both \cite{Gui-letter} and \cite{xianghao-robust} only considered
the bounded CSI error model. 
\item For the bounded CSI error, we formulate worst-case robust beamforming
design problems that minimize the transmit power subject to unit modulus
of the reflection beamforming and the worst-case QoS constraints with
imperfect CBIUT. The worst-case robust design can guarantee that the
achievable rate of each user is no less than its minimum rate requirement
for all possible channel error realizations. To address this non-convex
problem, S-procedure is firstly adopted to approximate the semi-infinite
inequality constraints. Then, under the alternate optimization (AO)
framework, the precoder is updated in an second-order cone programming
(SOCP) and the reflection beamforming is updated by using the penalty
convex-concave procedure (CCP). 
\item For the statistical CSI error model, we aim to minimize the transmit
power subject to unit-modulus constraints and the rate outage probability
constraints. Here, the rate outage probability constraints represent
the probability that the achievable rate of each user being below
its minimum rate requirment needs to be less than a predetermined
probability. By applying the Bernstein-Type Inequality, the safe approximation
of the rate outage probability is obtained to make the original problem
tractable. Then, the precoder and the reflection beamforming are optimized
by using the semidefinite relaxation (SDR) and penalty CCP techniques
respectively in an iterative manner. 
\item We demonstrate through numerical results that the robust beamforming
under the statistical CSI error model can achieve superior system
performance in terms of the minimum transmit power, convergence speed
and complexity, than that under the bounded CSI error model. In addition,
it is observed that the level of the CBIUT error plays an important
role in the IRS-aided systems. Specifically, when the CBIUT error
is small, the total transmit power decreases with the number of the
reflection elements due to the increased beamforming gain. However,
when the CBIUT error is large, the transmit power increases with the
number of the reflection elements due to the increased channel estimation
error. Hence, whether to deploy the IRS in wireless communication
systems depends on the level of the CBIUT error. 
\end{itemize}
\,\,\,\,\,\,\,The remainder of this paper is organized as follows.
Section II introduces the system model and the CSI error models. Worst-case
robust design problems are formulated and solved in Section III. Section
IV further investigates the outage constrained robust design problems.
Section V compares the computational complexity of the developed robust
design methods. Finally, Section VI and Section VII show the numerical
results and conclusions, respectively.

\noindent \textbf{Notations:} The following mathematical notations
and symbols are used throughout this paper. Vectors and matrices are
denoted by boldface lowercase letters and boldface uppercase letters,
respectively. The symbols $\mathbf{X}^{*}$, $\mathbf{X}^{\mathrm{T}}$,
$\mathbf{X}^{\mathrm{H}}$, and $||\mathbf{X}||_{F}$ denote the conjugate,
transpose, Hermitian (conjugate transpose), Frobenius norm of matrix
$\mathbf{X}$, respectively. The symbol $||\mathbf{x}||_{2}$ denotes
2-norm of vector $\mathbf{x}$. The symbols $\mathrm{Tr}\{\cdot\}$,
$\mathrm{Re}\{\cdot\}$, $|\cdot|$, $\lambda(\cdot)$, and $\angle\left(\cdot\right)$
denote the trace, real part, modulus, eigenvalue, and angle of a complex
number, respectively. $\mathrm{diag}(\mathbf{x})$ is a diagonal matrix
with the entries of $\mathbf{x}$ on its main diagonal. $[\mathbf{x}]_{m}$
means the $m^{\mathrm{th}}$ element of the vector $\mathbf{x}$.
The Kronecker product between two matrices $\mathbf{X}$ and $\mathbf{Y}$
is denoted by $\mathbf{X}\otimes\mathbf{Y}$. $\mathbf{X}\succeq\mathbf{Y}$means
that $\mathbf{X}-\mathbf{Y}$ is positive semidefinite. Additionally,
the symbol $\mathbb{C}$ denotes complex field, $\mathbb{R}$ represents
real field, and $j\triangleq\sqrt{-1}$ is the imaginary unit.

\section{System Model}

In this section, we first introduce the system model of the IRS-aided
MISO downlink communication system, and then discuss the channel uncertainty
scenarios as well as the CSI error models.

\subsection{Signal Transmission Model}

\begin{figure}
\centering \includegraphics[width=3.5in,height=2.5in]{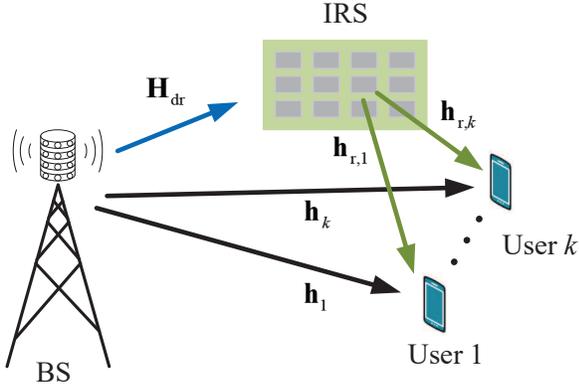}
\caption{An IRS-aided multi-user communication system.}
\label{system-model} 
\end{figure}

As shown in Fig. \ref{system-model}, we consider an IRS-aided MISO
broadcast (BC) communication system, which consists of one multi-antenna
BS, $K$ single-antenna users and one IRS. It is assumed that the
BS is equipped with $N$ active antennas, and transmits $K$ Gaussian
data symbols denoted by $\mathbf{s}=[s_{1},\cdots,s_{K}]^{\mathrm{T}}\in\mathbb{C}^{K\times1}$
to all the users, where $\mathbb{E}[\mathbf{s}\mathbf{s}^{\mathrm{H}}]=\mathbf{I}$.
IRS with $M$ programmable phase shifters is deployed to enhance the
system performance. Therefore, by defining the set of users as $\mathcal{K}=\{1,2,...,K\}$,
the received baseband signal of users is given by 
\begin{equation}
y_{k}=(\mathbf{h}_{k}^{\mathrm{H}}+\mathbf{h}_{\mathrm{r},k}^{\mathrm{H}}\mathbf{E}\mathbf{H_{\mathrm{dr}}}){\bf F}{\bf s}+n_{k},\forall k\in\mathcal{K}.\label{eq:received signal}
\end{equation}
Here, $\mathbf{F=}[{\bf \mathbf{f}}_{1},\cdots,{\bf \mathbf{f}}_{K}]\in\mathbb{C}^{N\times K}$
is the precoder matrix, in which ${\bf \mathbf{f}}_{k}$ is the precoding
vector associated with user $k$. Then, the transmit power at the
BS is $\mathbb{E}\{\mathrm{Tr}\left[{\bf F}{\bf s}\mathbf{s}^{\mathrm{H}}\mathbf{F}^{\mathrm{H}}\right]\}=||\mathbf{F}||_{F}^{2}$.
$n_{k}$ is the additive white Gaussian noise (AWGN) at user $k$,
with zero mean and noise variance $\sigma_{k}^{2}$, i.e., $n_{k}\sim\mathcal{CN}(0,\sigma_{k}^{2})$.
The reflection beamforming of the IRS is a diagonal matrix $\mathbf{E}=\sqrt{\iota}\mathrm{diag}(e_{1},\cdots,e_{M})\in\mathbb{C}^{M\times M}$,
of which has unit-modulus phase shifts, i.e., $|e_{m}|^{2}=1$. $0\leq\iota\leq1$
indicates the reflection efficiency and the power loss of reflection
operation usually comes from multiple reflections of signals. Here,
we assume that only the first-order reflection on the IRS is considered
and set $\iota=1$. It is assumed that the phase shifts of the IRS
are calculated by the BS and then fed back to the IRS controller through
dedicated feedback channels \cite{Pan2019intelleget,Pan2019multicell}.
In addition, the channel vectors spanning from the BS to user $k$
and from the IRS to user $k$ are denoted by $\mathbf{h}_{k}\in\mathbb{C}^{N\times1}$
and $\mathbf{h}_{\mathrm{r},k}\in\mathbb{C}^{M\times1}$, respectively.
The channel matrix between the BS and the IRS is represented by $\mathbf{H_{\mathrm{dr}}}\in\mathbb{C}^{M\times N}$.

Denote by $\mathbf{G}_{k}=\mathrm{diag}(\mathbf{h}_{\mathrm{r},k}^{\mathrm{H}})\mathbf{H_{\mathrm{dr}}}$
the cascaded channel from the BS to user $k$ via the IRS, by $\mathbf{e}=[e_{1},\cdots,e_{M}]^{\mathrm{T}}\in\mathbb{C}^{M\times1}$
the vector containing diagonal elements of matrix $\mathbf{E}$, and
by $\beta_{k}=||(\mathbf{h}_{k}^{\mathrm{H}}+\mathbf{e}^{\mathrm{H}}\mathbf{G}_{k}){\bf F}_{-k}||_{2}^{2}+\sigma_{k}^{2}$
the interference-plus-noises (INs) power of user $k$, where ${\bf F}_{-k}=[{\bf \mathbf{f}}_{1},\cdots,{\bf \mathbf{f}}_{k-1},{\bf \mathbf{f}}_{k+1},\cdots,{\bf \mathbf{f}}_{K}]$.
Then, the achievable data rate (bit/s/Hz) at user $k$ is given by
\begin{align}
 & \mathcal{R}_{k}\left(\mathbf{F},\mathbf{e}\right)=\log_{2}\left(1+\frac{1}{\beta_{k}}\left|\left(\mathbf{h}_{k}^{\mathrm{H}}+\mathbf{e}^{\mathrm{H}}\mathbf{G}_{k}\right){\bf f}_{k}\right|^{2}\right).\label{eq:rate-1}
\end{align}

\subsection{Two Scenarios and CSI Error Models}

In the IRS-aided communication system, there are two types of channels:
the direct channel $\mathbf{h}_{k}$, and the cascaded BS-IRS-user
channel $\mathbf{G}_{k}$. The system performance of the IRS-aided
communication system is highly affected by the accuracy of the direct
channel state information at the transmitter (DCSIT) and the CBIUT.
In the following, we first introduce two scenarios of the channel
uncertainties and then two types of CSI error models.

\textit{1) Scenario 1: Partial Channel Uncertainty (PCU)}

In IRS-aided communications, the CBIUT is much more challenging to
obtain than the DCSIT due to the passive features of the IRS. Hence,
in this scenario, we assume that the DCSIT is perfect, while the CBIUT
is imperfect. The CBIUT can be represented as 
\begin{equation}
\mathbf{G}_{k}=\widehat{\mathbf{G}}_{k}+\bigtriangleup\mathbf{G}_{k},\forall k\in\mathcal{K},\label{eq:G-channel}
\end{equation}
where $\widehat{\mathbf{G}}_{k}$ is the estimated cascaded CSI known
at the BS, $\bigtriangleup\mathbf{G}_{k}$ is the unknown CBIUT error.

\textit{2) Scenario 2: Full Channel Uncertainty (FCU)}

In complex electromagnetic environment, the accurate DCSIT is also
challenging to obtain. In this scenario, we assume both the DCSIT
and the CBIUT are imperfect. In addition to the CBIUT error model
in (\ref{eq:G-channel}), the direct channel is expressed as 
\begin{equation}
\mathbf{h}_{k}=\widehat{\mathbf{h}}_{k}+\bigtriangleup\mathbf{h}_{k},\forall k\in\mathcal{K},\label{eq:H-channel}
\end{equation}
where $\widehat{\mathbf{h}}_{k}$ is the estimated DCSIT known at
the BS and $\bigtriangleup\mathbf{h}_{k}$ is the unknown DCSIT error.

In this work, we investigate two types of robust beamforming design
for IRS-aided MISO communication systems depending on the CSI error
models.

\textit{1) Error model 1: Bounded CSI error model}

Specifically, one is the worst-case robust beamforming design subject
to the bounded CSI error model, i.e., 
\begin{equation}
\left\Vert \bigtriangleup\mathbf{G}_{k}\right\Vert _{F}\leq\xi_{\mathrm{g},k},\left\Vert \bigtriangleup\mathbf{h}_{k}\right\Vert _{2}\leq\xi_{\mathrm{h},k},\forall k\in\mathcal{K},\label{eq:bounded-error}
\end{equation}
where $\xi_{\mathrm{g},k}$ and $\xi_{\mathrm{h},k}$ are the radii
of the uncertainty regions known at the BS. This CSI error model characterizes
the channel quantization error which naturally belongs to a bounded
region \cite{bounded-channel}. For example, in the frequency division
duplex (FDD) setting, the receiver estimates the downlink channel
and then feeds the rate-limited quantized CSI back to the transmitter.
Then, the acquired CSI is plagued by quantization errors.

\textit{2) Error model 2: Statistical CSI error model}

The other is the outage-constrained robust beamforming design associated
with the statistical CSI error model, in which each CSI error vector
is assumed to follow the circularly symmetric complex Gaussian (CSCG)
distribution, i.e., \begin{subequations}\label{Pro:statistic-error}
\begin{align}
\mathrm{vec}(\bigtriangleup\mathbf{G}_{k}) & \sim\mathcal{CN}(\mathbf{0},\boldsymbol{\Sigma}_{\mathrm{g},k}),\boldsymbol{\Sigma}_{\mathrm{g},k}\succeq\mathbf{0},\forall k\in\mathcal{K},\\
\bigtriangleup\mathbf{h}_{k} & \sim\mathcal{CN}(\mathbf{0},\boldsymbol{\Sigma}_{\mathrm{h},k}),\boldsymbol{\Sigma}_{\mathrm{h},k}\succeq\mathbf{0},\forall k\in\mathcal{K},
\end{align}
\end{subequations}where $\boldsymbol{\Sigma}_{\mathrm{g},k}\in\mathbb{C}^{MN\times MN}$
and $\boldsymbol{\Sigma}_{\mathrm{h},k}\in\mathbb{C}^{N\times N}$
are positive semidefinite error covariance matrices. In this case,
the CSI imperfection is caused by the channel estimation error \cite{jun-channel}.
For example, in the time division duplex (TDD) setting, noise and
limited training will cause the uplink channel estimation error. The
conventional MMSE method is generally adopted to estimate the cascaded
channel, and thus the channel estimation generally follows the CSCG
distribution.

In the following, we first consider the first type of robust beamforming
design based on the bounded CSI error model. Then, we deal with the
second one based on the statistical CSI error model.

\section{Worst-case robust beamforming design}

In this section, the worst-case robust beamforming design is considered
under the bounded CSI error model. We aim to minimize the total transmit
power of the BS by the joint design of the precoder matrix $\mathbf{F}$
and reflection beamforming vector $\mathbf{e}$ under the unit-modulus
constraints and the worst-case QoS constraints, i.e., ensuring the
achievable rate of each user to be above a threshold for all possible
channel error realizations. In order to solve the non-convex robust
design problem with semi-infinite inequality constraints and coupled
variables, an AO algorithm is proposed based on S-Procedure, SOCP
and penalty CCP \cite{PCCP-boyd}.

First, two useful lemmas about multiple complex valued uncertainties
are formally introduced as follows, which will be used in the later
derivations.

\begin{lemma}\label{S-procedure}(General S-Procedure \cite{boy-S-procedure})
Define the quadratic functions of the variable $\mathbf{x}\in\mathbb{C}^{n\times1}$:
\[
f_{i}(\mathbf{x})=\mathbf{x}^{\mathrm{H}}\mathbf{W}_{i}\mathbf{x}+2\mathrm{Re}\left\{ \mathbf{w}_{i}^{\mathrm{H}}\mathbf{x}\right\} +w_{i},\thinspace\thinspace i=0,...,P,
\]
where $\mathbf{W}_{i}=\mathbf{W}_{i}^{\mathrm{H}}$. The condition
$\left\{ f_{i}(\mathbf{x})\geq0\right\} _{i=1}^{P}\Rightarrow f_{0}(\mathbf{x})\geq0$
holds if and only if there exist $\forall i,\varpi_{i}\geq0$ such
that 
\[
\left[\begin{array}{cc}
\mathbf{W}_{0} & \mathbf{w}_{0}\\
\mathbf{w}_{0}^{\mathrm{H}} & w_{0}
\end{array}\right]-\sum_{i=1}^{P}\varpi_{i}\left[\begin{array}{cc}
\mathbf{W}_{i} & \mathbf{w}_{i}\\
\mathbf{w}_{i}^{\mathrm{H}} & w_{i}
\end{array}\right]\succeq\mathbf{0}.
\]

\end{lemma}

\begin{lemma}\label{sign-definiteness}(General sign-definiteness
\cite{Petersen-lemma}) For a given set of matrices $\mathbf{W}=\mathbf{W}^{\mathrm{H}}$,
$\{\mathbf{Y}_{i},\mathbf{Z}_{i}\}_{i=1}^{P}$, the following linear
matrix inequality (LMI) satisfies 
\[
\mathbf{W}\succeq\sum_{i=1}^{P}\left(\mathbf{Y}_{i}^{\mathrm{H}}\mathbf{X}_{i}\mathbf{Z}_{i}+\mathbf{Z}_{i}^{\mathrm{H}}\mathbf{X}_{i}^{\mathrm{H}}\mathbf{Y}_{i}\right),\forall i,||\mathbf{X}_{i}||_{F}\leq\xi_{i},
\]
if and only if there exist real numbers $\forall i,\mu_{i}\geq0$
such that 
\[
\left[\begin{array}{cccc}
\mathbf{W}-\sum_{i=1}^{P}\mu_{i}\mathbf{Z}_{i}^{\mathrm{H}}\mathbf{Z}_{i} & -\xi_{1}\mathbf{Y}_{1}^{\mathrm{H}} & \cdots & -\xi_{P}\mathbf{Y}_{P}^{\mathrm{H}}\\
-\xi_{1}\mathbf{Y}_{1} & \mu_{1}\mathbf{I} & \cdots & \mathbf{0}\\
\vdots & \vdots & \ddots & \vdots\\
-\xi_{P}\mathbf{Y}_{P} & \mathbf{0} & \cdots & \mu_{P}\mathbf{I}
\end{array}\right]\succeq\mathbf{0}.
\]

\end{lemma}

It is noted that Lemma \ref{sign-definiteness} can be proved by applying
Lemma \ref{S-procedure} and the detailed proof is given in \cite{Gharavol2013TSP}.

\subsection{Scenario 1: Partial Channel Uncertainty}

In this subsection, we design the robust beamforming for the IRS-aided
communication system under Scenario 1 with perfect DCSIT and imperfect
CBIUT. This problem is simpler than the one with full channel uncertainty
and the algorithm developed for Scenario 1 has lower complexity than
that for Scenario 2. Mathematically, let $\mathcal{E}_{k}^{partial}\triangleq\{\forall\left\Vert \bigtriangleup\mathbf{G}_{k}\right\Vert _{F}\leq\xi_{\mathrm{g},k}\}$
and denote by $\mathcal{M}=\{1,2,...,M\}$ the set of reflection elements,
the worst-case transmit power minimization problem is formulated as
\begin{subequations}\label{Pro:min-power-worst-partial} 
\begin{align}
\mathop{\min}\limits _{\mathbf{F},\mathbf{e}} & \thinspace\thinspace||\mathbf{F}||_{F}^{2}\label{eq:min-power-obj-1}\\
\textrm{s.t.} & \thinspace\thinspace\mathcal{R}_{k}\left(\mathbf{F},\mathbf{e}\right)\geq R_{k},\mathcal{E}_{k}^{partial},\forall k\in\mathcal{K}\label{eq:min-power-cons1-1}\\
 & \thinspace\thinspace|e_{m}|^{2}=1,\forall m\in\mathcal{M}.\label{eq:min-power-cons2}
\end{align}
\end{subequations}Here, $R_{k}$ is the target rate of user $k$.
Constraints (\ref{eq:min-power-cons1-1}) are the worst-case QoS requirements
for the users, while constraints (\ref{eq:min-power-cons2}) correspond
to the unit-modulus requirements of the reflection elements at the
IRS.

To start with, the non-convexity of constraints (\ref{eq:min-power-cons1-1})
can be addressed by firstly treating the INs power $\boldsymbol{\beta}=[\beta_{1},...,\beta_{K}]^{\mathrm{T}}$
as auxiliary variables. Hence, constraints (\ref{eq:min-power-cons1-1})
are reformulated as 
\begin{align}
 & \left|\left(\mathbf{h}_{k}^{\mathrm{H}}+\mathbf{e}^{\mathrm{H}}\mathbf{G}_{k}\right){\bf f}_{k}\right|^{2}\geq\beta_{k}(2^{R_{k}}-1),\mathcal{E}_{k}^{partial},\forall k\in\mathcal{K},\label{eq:single-1}\\
 & \left\Vert \left(\mathbf{h}_{k}^{\mathrm{H}}+\mathbf{e}^{\mathrm{H}}\mathbf{G}_{k}\right){\bf F}_{-k}\right\Vert _{2}^{2}+\sigma_{k}^{2}\leq\beta_{k},\mathcal{E}_{k}^{partial},\forall k\in\mathcal{K}.\label{eq:IN-1}
\end{align}
Constraints (\ref{eq:single-1}) and (\ref{eq:IN-1}) are termed as
the worst-case useful signal power constraints and the worst-case
INs power constraints, respectively.

Then, the non-convex semi-infinite inequality constraints (\ref{eq:single-1})
are handled by firstly approximating the non-convex parts and then
dealing with the semi-infinite inequalities by using the S-Procedure.
Specifically, the following lemma shows the linear approximation of
the useful signal power in (\ref{eq:single-1}).

\begin{lemma}\label{lower-bound} Substituting $\mathbf{G}_{k}=\widehat{\mathbf{G}}_{k}+\bigtriangleup\mathbf{G}_{k}$
into the useful signal power in (\ref{eq:single-1}) and let $\mathbf{f}_{k}^{(n)}$
and $\mathbf{e}^{(n)}$ be the optimal solutions obtained at iteration
$n$, then $|[\mathbf{h}_{k}^{\mathrm{H}}+\mathbf{e}^{\mathrm{H}}(\widehat{\mathbf{G}}_{k}+\bigtriangleup\mathbf{G}_{k})]{\bf f}_{k}|^{2}$
is linearly approximated by its lower bound at ($\mathbf{f}_{k}^{(n)}$,
$\mathbf{e}^{(n)}$) as follows 
\begin{equation}
\mathrm{vec}^{\mathrm{T}}(\mathbf{\bigtriangleup\mathbf{G}}_{k})\mathbf{A}_{k}\mathrm{vec}(\mathbf{\bigtriangleup\mathbf{G}}_{k}^{*})+2\mathrm{Re}\left\{ \mathbf{a}_{k}^{\mathrm{T}}\mathrm{vec}(\mathbf{\bigtriangleup\mathbf{G}}_{k}^{*})\right\} +a_{k},\label{eq:lower-bound-1}
\end{equation}
where 
\begin{align*}
\mathbf{A}_{k} & =\mathbf{f}_{k}{\bf f}_{k}^{(n),\mathrm{H}}\otimes\mathbf{e}^{*}\mathbf{e}^{(n),\mathrm{T}}+\mathbf{f}_{k}^{(n)}{\bf f}_{k}^{\mathrm{H}}\otimes\mathbf{e}^{(n),*}\mathbf{e}^{\mathrm{T}}\\
 & \thinspace\thinspace\thinspace\thinspace\thinspace\thinspace\thinspace-(\mathbf{f}_{k}^{(n)}{\bf f}_{k}^{(n),\mathrm{H}}\otimes\mathbf{e}^{(n),*}\mathbf{e}^{(n),\mathrm{T}}),\\
\mathbf{a}_{k} & =\mathrm{vec}(\mathbf{e}\left(\mathbf{h}_{k}^{\mathrm{H}}+\mathbf{e}^{(n),\mathrm{H}}\mathbf{\widehat{G}}_{k}\right){\bf f}_{k}^{(n)}\mathbf{f}_{k}^{\mathrm{H}})\\
 & \thinspace\thinspace\thinspace\thinspace\thinspace\thinspace\thinspace+\mathrm{vec}(\mathbf{e}^{(n)}\left(\mathbf{h}_{k}^{\mathrm{H}}+\mathbf{e}^{\mathrm{H}}\mathbf{\widehat{G}}_{k}\right){\bf f}_{k}\mathbf{f}_{k}^{(n),\mathrm{H}})\\
 & \thinspace\thinspace\thinspace\thinspace\thinspace\thinspace\thinspace-\mathrm{vec}(\mathbf{e}^{(n)}\left(\mathbf{h}_{k}^{\mathrm{H}}+\mathbf{e}^{(n),\mathrm{H}}\mathbf{\widehat{G}}_{k}\right){\bf f}_{k}^{(n)}\mathbf{f}_{k}^{(n),\mathrm{H}}),\\
a_{k} & =2\mathrm{Re}\left\{ \left(\mathbf{h}_{k}^{\mathrm{H}}+\mathbf{e}^{(n),\mathrm{H}}\mathbf{\widehat{G}}_{k}\right){\bf f}_{k}^{(n)}\mathbf{f}_{k}^{\mathrm{H}}\left(\mathbf{h}_{k}+\mathbf{\widehat{G}}_{k}^{\mathrm{H}}\mathbf{e}\right)\right\} \\
 & \thinspace\thinspace\thinspace\thinspace\thinspace\thinspace\thinspace-\left(\mathbf{h}_{k}^{\mathrm{H}}+\mathbf{e}^{(n),\mathrm{H}}\mathbf{\widehat{G}}_{k}\right){\bf f}_{k}^{(n)}\mathbf{f}_{k}^{(n),\mathrm{H}}\left(\mathbf{h}_{k}+\mathbf{\widehat{G}}_{k}^{\mathrm{H}}\mathbf{e}^{(n)}\right).
\end{align*}
\end{lemma}

\textbf{\textit{Proof: }}Please refer to Appendix \ref{subsec:The-proof-of-1}.\hspace{3cm}$\blacksquare$

By replacing the useful signal power in (\ref{eq:single-1}) with
its linear approximation (\ref{eq:lower-bound-1}), constraints (\ref{eq:single-1})
are reformulated as 
\begin{align}
 & \mathrm{vec}^{\mathrm{T}}(\mathbf{\bigtriangleup\mathbf{G}}_{k})\mathbf{A}_{k}\mathrm{vec}(\mathbf{\bigtriangleup\mathbf{G}}_{k}^{*})+2\mathrm{Re}\left\{ \mathbf{a}_{k}^{\mathrm{T}}\mathrm{vec}(\mathbf{\bigtriangleup\mathbf{G}}_{k}^{*})\right\} +a_{k}\nonumber \\
 & \geq\beta_{k}(2^{R_{k}}-1),\mathcal{E}_{k}^{partial},\forall k\in\mathcal{K}.\label{eq:signal-1}
\end{align}

Lemma \ref{S-procedure} is then used to tackle the CSI uncertainty
in the above constraints. Specifically, constraint corresponding to
each user in (\ref{eq:signal-1}) can be recast by setting the parameters
in Lemma \ref{S-procedure} as follows 
\begin{align*}
 & P=1,\thinspace\thinspace\mathbf{W}_{0}=\mathbf{A}_{k},\thinspace\thinspace\mathbf{w}_{0}=\mathbf{a}_{k},\thinspace\thinspace w_{0}=a_{k}-\beta_{k}(2^{R_{k}}-1),\\
 & \mathbf{x}=\mathrm{vec}(\mathbf{\bigtriangleup\mathbf{G}}_{k}^{*}),\thinspace\thinspace\thinspace\thinspace\thinspace\thinspace\thinspace\thinspace\thinspace\mathbf{W}_{1}=-\mathbf{I},\thinspace\thinspace\thinspace w_{1}=\xi_{k}^{2}.
\end{align*}

Then, (\ref{eq:signal-1}) is transformed into the following equivalent
LMIs as 
\begin{align}
 & \left[\begin{array}{cc}
\varpi_{\mathrm{g},k}\mathbf{I}_{MN}+\mathbf{A}_{k} & \mathbf{a}_{k}\\
\mathbf{a}_{k}^{\mathrm{T}} & C_{k}^{partial}
\end{array}\right]\succeq\mathbf{0},\forall k\in\mathcal{K},\label{eq:LMI-S-partial}
\end{align}
where $\boldsymbol{\varpi}_{\mathrm{g}}=[\varpi_{\mathrm{g},1},...,\varpi_{\mathrm{g},K}]^{\mathrm{T}}\geq0$
are slack variables and $C_{k}^{partial}=a_{k}-\beta_{k}(2^{R_{k}}-1)-\varpi_{\mathrm{g},k}\xi_{k}^{2}$.

Next, we consider the uncertainty in $\{\bigtriangleup\mathbf{G}_{k}\}_{\forall k\in\mathcal{K}}$
of (\ref{eq:IN-1}). Specifically, we firstly adopt Schur's complement
Lemma \cite{book-convex} to equivalently recast the INs power inequalities
in (\ref{eq:IN-1}) into matrix inequalities as follows 
\begin{align}
 & \left[\begin{array}{cc}
\beta_{k}-\sigma_{k}^{2} & \mathbf{t}_{k}^{\mathrm{H}}\\
\mathbf{t}_{k} & \mathbf{I}
\end{array}\right]\succeq\mathbf{0},\forall k\in\mathcal{K},\label{eq:IN-LMI-1-1}
\end{align}
where $\mathbf{t}_{k}=((\mathbf{h}_{k}^{\mathrm{H}}+\mathbf{e}^{\mathrm{H}}\mathbf{G}_{k}){\bf F}_{-k})^{\mathrm{H}}$.
By using $\mathbf{G}_{k}=\widehat{\mathbf{G}}_{k}+\bigtriangleup\mathbf{G}_{k}$,
(\ref{eq:IN-LMI-1-1}) is then rewritten as 
\begin{align}
\left[\begin{array}{cc}
\beta_{k}-\sigma_{k}^{2} & \widehat{\mathbf{t}}_{k}^{\mathrm{H}}\\
\widehat{\mathbf{t}}_{k} & \mathbf{I}
\end{array}\right]\succeq & -\left[\begin{array}{c}
\mathbf{0}\\
{\bf F}_{-k}^{\mathrm{H}}
\end{array}\right]\triangle\mathbf{G}_{k}^{\mathrm{H}}\left[\begin{array}{cc}
\mathbf{e} & \mathbf{0}\end{array}\right]\nonumber \\
-\left[\begin{array}{c}
\mathbf{e}^{\mathrm{H}}\\
\mathbf{0}
\end{array}\right] & \triangle\mathbf{G}_{k}\left[\begin{array}{cc}
\mathbf{0} & {\bf F}_{-k}\end{array}\right],\forall k\in\mathcal{K},\label{eq:LMI-IN-2}
\end{align}
where $\widehat{\mathbf{t}}_{k}=((\mathbf{h}_{k}^{\mathrm{H}}+\mathbf{e}^{\mathrm{H}}\widehat{\mathbf{G}}_{k}){\bf F}_{-k})^{\mathrm{H}}$.

In order to use Lemma \ref{sign-definiteness}, we choose the following
parameters (It is noted that the subscript $i$ in Lemma \ref{sign-definiteness}
has been ignored since $P=1$.) for each constraint in (\ref{eq:LMI-IN-2})
as 
\begin{align*}
\mathbf{W} & =\left[\begin{array}{cc}
\beta_{k}-\sigma_{k}^{2} & \widehat{\mathbf{t}}_{k}^{\mathrm{H}}\\
\widehat{\mathbf{t}}_{k} & \mathbf{I}
\end{array}\right],\mathbf{Y}=-\left[\begin{array}{cc}
\mathbf{0} & {\bf F}_{-k}\end{array}\right],\\
\mathbf{Z} & =\left[\begin{array}{cc}
\mathbf{e} & \mathbf{0}\end{array}\right],\thinspace\thinspace\thinspace\thinspace\thinspace\thinspace\thinspace\thinspace\thinspace\thinspace\thinspace\thinspace\thinspace\thinspace\thinspace\thinspace\thinspace\thinspace\thinspace\thinspace\thinspace\mathbf{X}=\triangle\mathbf{G}_{k}^{\mathrm{H}}.
\end{align*}

Then, the equivalent LMIs of the worst-case INs power constraints
(\ref{eq:IN-1}) are given by 
\begin{equation}
\left[\begin{array}{ccc}
\beta_{k}-\sigma_{k}^{2}-\mu_{\mathrm{g},k}M & \widehat{\mathbf{t}}_{k}^{\mathrm{H}} & \mathbf{0}_{1\times N}\\
\widehat{\mathbf{t}}_{k} & \mathbf{I}_{(K-1)} & \xi_{\mathrm{g},k}{\bf F}_{-k}^{\mathrm{H}}\\
\mathbf{0}_{N\times1} & \xi_{\mathrm{g},k}{\bf F}_{-k} & \mu_{\mathrm{g},k}\mathbf{I}_{N}
\end{array}\right]\succeq\mathbf{0},\forall k\in\mathcal{K},\label{eq:LMI-IN-1-1}
\end{equation}
where $\boldsymbol{\mu}_{\mathrm{g}}=[\mu_{\mathrm{g},1},...,\mu_{\mathrm{g},K}]^{\mathrm{T}}\geq0$
are slack variables.

Based on the above discussions, Problem (\ref{Pro:min-power-worst-partial})
is approximately rewritten as \begin{subequations}\label{Pro:min-power-worst-partial-2}
\begin{align}
\mathop{\min}\limits _{\mathbf{F},\mathbf{e},\boldsymbol{\beta},\boldsymbol{\varpi}_{\mathrm{g}},\boldsymbol{\mu}_{\mathrm{g}}} & \thinspace\thinspace||\mathbf{F}||_{F}^{2}\label{eq:obj-2}\\
{\rm s.t.} & \thinspace\thinspace(\ref{eq:LMI-S-partial}),(\ref{eq:LMI-IN-1-1}),(\ref{eq:min-power-cons2}),\\
 & \thinspace\thinspace\boldsymbol{\varpi}_{\mathrm{g}}\geq0,\boldsymbol{\mu}_{\mathrm{g}}\geq0.\label{eq:omiga}
\end{align}
\end{subequations}This problem is still non-convex and difficult
to optimize $\mathbf{F}$ and $\mathbf{e}$ simultaneously since $\mathbf{F}$
and $\mathbf{e}$ are coupled in $\mathbf{A}_{k}$, $\mathbf{a}_{k}$
and $\widehat{\mathbf{t}}_{k}$. In the following, we adopt the AO
method to optimize $\mathbf{F}$ and $\mathbf{e}$ sequentially in
an iterative manner. In particular, we minimize the transmit power
by first fixing the reflection beamforming $\mathbf{e}$ so that the
problem reduces to a convex one with respect to $\mathbf{F}$. CVX
tool \cite{CVX2018} is adopted to solve the resulting convex problem.
Precoder $\mathbf{F}$ is then fixed and the resulting non-convex
problem of $\mathbf{e}$ is handled under the penalty CCP method.
Specifically, for given $\mathbf{e}$, the subproblem of $\mathbf{F}$
is given by \begin{subequations}\label{Pro:min-power-worst-partial-f}
\begin{align}
\mathbf{F}^{(n+1)}=\mathrm{arg}\mathop{\min}\limits _{\mathbf{F},\boldsymbol{\beta},\boldsymbol{\varpi}_{\mathrm{g}},\boldsymbol{\mu}_{\mathrm{g}}} & \thinspace\thinspace||\mathbf{F}||_{F}^{2}\label{eq:obj-3}\\
{\rm s.t.} & \thinspace\thinspace(\ref{eq:LMI-S-partial}),(\ref{eq:LMI-IN-1-1}),(\ref{eq:omiga}),
\end{align}
\end{subequations}where $\mathbf{F}^{(n+1)}$ is the optimal solution
obtained in the $(n+1)$-th iteration. Problem (\ref{Pro:min-power-worst-partial-f})
is a semidefinite program (SDP) and can be solved by the CVX tool.

Then, for given $\mathbf{F}$, the subproblem of $\mathbf{e}$ is
a feasibility-check problem. According to \cite{qingqing2019,Gui-letter}
and in order to improve the converged solution in the optimization
of $\mathbf{e}$, the useful signal power inequalities in (\ref{eq:single-1})
are modified by introducing slack variables $\boldsymbol{\alpha}=[\alpha_{1},...,\alpha_{K}]^{\mathrm{T}}\geq0$
and recast as 
\begin{equation}
\left|\left(\mathbf{h}_{k}^{\mathrm{H}}+\mathbf{e}^{\mathrm{H}}\mathbf{G}_{k}\right){\bf f}_{k}\right|^{2}\geq\beta_{k}(2^{R_{k}}-1)+\alpha_{k},\forall k\in\mathcal{K}.\label{eq:SINR-slac}
\end{equation}
Subsequently, the LMIs (\ref{eq:LMI-S-partial}) are modified as 
\begin{align}
 & \left[\begin{array}{cc}
\varpi_{\mathrm{g},k}\mathbf{I}_{MN}+\mathbf{A}_{k} & \mathbf{a}_{k}\\
\mathbf{a}_{k}^{\mathrm{T}} & C_{k}^{partial}-\alpha_{k}
\end{array}\right]\succeq\mathbf{0},\forall k\in\mathcal{K}.\label{eq:LMI-S-partial-1}
\end{align}
In addition, we note that only the submatrix of $K\times K$ in the
upper left corner of (\ref{eq:LMI-IN-1-1}) depends on $\mathbf{e}$,
so the dimension of the LMIs (\ref{eq:LMI-IN-1-1}) can be reduced
from $(K+N)\times(K+N)$ to $K\times K$ as 
\begin{equation}
\left[\begin{array}{cc}
\beta_{k}-\sigma_{k}^{2}-\mu_{\mathrm{g},k}M & \widehat{\mathbf{t}}_{k}^{\mathrm{H}}\\
\widehat{\mathbf{t}}_{k} & \mathbf{I}_{(K-1)}
\end{array}\right]\succeq\mathbf{0},\forall k\in\mathcal{K}.\label{eq:LMI-IN-partial-1}
\end{equation}
Combining (\ref{eq:LMI-S-partial-1}) and (\ref{eq:LMI-IN-partial-1}),
the sub-problem of $\mathbf{e}$ can be formulated as \begin{subequations}\label{Pro:min-power-4}
\begin{align}
\max\limits _{\boldsymbol{\alpha},\mathbf{e},\boldsymbol{\beta},\boldsymbol{\varpi}_{\mathrm{g}},\boldsymbol{\mu}_{\mathrm{g}}} & \thinspace\thinspace\sum_{k=1}^{K}\alpha_{k}\label{eq:obj-4}\\
{\rm s.t.} & \thinspace\thinspace(\ref{eq:LMI-S-partial-1}),(\ref{eq:LMI-IN-partial-1}),(\ref{eq:min-power-cons2}),(\ref{eq:omiga}),\\
 & \thinspace\thinspace\boldsymbol{\alpha}\geq0.\label{eq:SINR residual}
\end{align}
\end{subequations} Note that the solution of Problem (\ref{Pro:min-power-4})
can yield a lower objective value compared with Problem (\ref{Pro:min-power-worst-partial-f}),
the explanation of which can be found in \cite{qingqing2019}.

We note that the above problem is still non-convex due to the unit-modulus
constraints. As in our previous work \cite{Gui-letter}, we here adopt
the penalty CCP \cite{PCCP-boyd} to deal with the non-convex constraints.
Following the penalty CCP framwork, the constraints (\ref{eq:min-power-cons2})
are firstly equivalently rewritten as $1\leq|e_{m}|^{2}\leq1,\forall m\in\mathcal{M}$.
The non-convex parts of the resulting constraints are then linearized
by $|e_{m}^{[t]}|^{2}-2\mathrm{Re}(e_{m}^{*}e_{m}^{[t]})\leq-1,\forall m\in\mathcal{M}$,
at fixed $e_{m}^{[t]}$. We finally have the following convex subproblem
of $\mathbf{e}$ as \begin{subequations}\label{Pro:min-power-5}
\begin{align}
\max\limits _{{\scriptstyle {\mathbf{e},\boldsymbol{\alpha},\mathbf{b},\hfill\atop {\scriptstyle \boldsymbol{\beta},\boldsymbol{\varpi}_{\mathrm{g}},\boldsymbol{\mu}_{\mathrm{g}}\hfill}}}} & \thinspace\thinspace\sum_{k=1}^{K}\alpha_{k}-\lambda^{[t]}\sum_{m=1}^{2M}b_{m}\label{eq:obj-4-1}\\
{\rm s.t.} & \textrm{\thinspace\thinspace(\ref{eq:LMI-S-partial-1}),(\ref{eq:LMI-IN-partial-1}),(\ref{eq:omiga}),(\ref{eq:SINR residual})},\\
 & \thinspace\thinspace|e_{m}^{[t]}|^{2}-2\mathrm{Re}(e_{m}^{*}e_{m}^{[t]})\leq b_{m}-1,\forall m\in\mathcal{M}\label{eq:unit-3-1}\\
 & \thinspace\thinspace|e_{m}|^{2}\leq1+b_{M+m},\forall m\in\mathcal{M}\\
 & \thinspace\thinspace\mathbf{b}\geq0,
\end{align}
\end{subequations} where $\mathbf{b}=[b_{1},...,b_{2M}]^{\mathrm{T}}$are
slack variables imposed over the equivalent linear constraints of
the unit-modulus constraints, and $||\mathbf{b}||_{1}$ is the penalty
term in the objective function. $||\mathbf{b}||_{1}$ is scaled by
the regularization factor $\lambda^{[t]}$ to control the feasibility
of the constraints.

Problem (\ref{Pro:min-power-5}) is an SDP and can be solved by the
CVX tool. The steps of finding a feasible solution of $\mathbf{e}$
to Problem (\ref{Pro:min-power-4}) is summarized in Algorithm \ref{Algorithm-analog}.
We remark that: \textit{a) }When $\chi$ is sufficiently low, constraints
(\ref{eq:min-power-cons2}) in the original Problem (\ref{Pro:min-power-4})
is guaranteed by $||\mathbf{b}||_{1}\leq\chi$; \textit{b) }The maximum
value $\lambda_{max}$ is imposed to avoid a numerical problem, that
is, a feasible solution satisfying $||\mathbf{b}||_{1}\leq\chi$ may
not be found when the iteration converges to the stopping criteria
$||\mathbf{e}^{[t]}-\mathbf{e}^{[t-1]}||_{1}\leq\nu$ with the increase
of $\lambda^{[t]}$; \textit{c) }Stopping criteria $||\mathbf{e}^{[t]}-\mathbf{e}^{[t-1]}||_{1}\leq\nu$
controls the convergence of Algorithm \ref{Algorithm-analog}; \textit{d)}
As mentioned in \cite{PCCP-boyd}, a feasible solution to Problem
(\ref{Pro:min-power-5}) may not be feasible for Problem (\ref{Pro:min-power-4}).
Hence, the feasibility of Problem (\ref{Pro:min-power-4}) is guaranteed
by imposing a maximum number of iterations $T_{max}$ and, in case
it is reached, we restart the iteration based on a new initial point.

\begin{algorithm}
\caption{Penalty CCP optimization for reflection beamforming optimization}
\label{Algorithm-analog} \begin{algorithmic}[1] \REQUIRE Initialize
$\mathbf{e}^{[0]}$, $\gamma^{[0]}>1$, and set $t=0$.

\REPEAT

\IF {$t<T_{max}$ }

\STATE Update $\mathbf{e}^{[t+1]}$ from Problem (\ref{Pro:min-power-5});

\STATE $\lambda^{[t+1]}=\min\{\gamma\lambda^{[t]},\lambda_{max}\}$;

\STATE $t=t+1$;

\ELSE

\STATE Initialize with a new random $\mathbf{e}^{[0]}$, set up $\gamma^{[0]}>1$
again, and set $t=0$.

\ENDIF

\UNTIL $||\mathbf{b}||_{1}\leq\chi$ and $||\mathbf{e}^{[t]}-\mathbf{e}^{[t-1]}||_{1}\leq\nu$.

\STATE Output $\mathbf{e}^{(n+1)}=\mathbf{e}^{[t]}$.

\end{algorithmic} 
\end{algorithm}

Finally, under the AO framework, Problem (\ref{Pro:min-power-worst-partial-2})
is solved by solving Problems (\ref{Pro:min-power-worst-partial-f})
and (\ref{Pro:min-power-4}) in an iterative manner. We remark that
the fixed point $\mathbf{e}^{[t]}$ in constraint (\ref{eq:unit-3-1})
is updated iteratively in Algorithm \ref{Algorithm-analog}, which
is the same as $\lambda^{[t]}$. While fixed point $\mathbf{e}^{(n)}$
in constraint (\ref{eq:LMI-S-partial-1}) is updated iteratively in
the outer AO framework.

\subsection{Scenario 2: Full Channel Uncertainty}

In this subsection, we extend the robust beamforming design in the
previous subsection to the case where both the DCSIT and CBIUT are
imperfect. By considering the full channel uncertainty in (\ref{eq:G-channel})
and (\ref{eq:H-channel}) and denoting $\mathcal{E}_{k}^{full}\triangleq\{\forall||\bigtriangleup\mathbf{h}_{k}||_{2}\leq\xi_{\mathrm{h},k},\forall||\bigtriangleup\mathbf{G}_{k}||_{F}\leq\xi_{\mathrm{g},k}\}$,
constraints (\ref{eq:min-power-cons1-1}) can be extended to 
\begin{equation}
\mathcal{R}_{k}\left(\mathbf{F},\mathbf{e}\right)\geq R_{k},\mathcal{E}_{k}^{full},\forall k\in\mathcal{K},\label{eq:rate-full-constraint}
\end{equation}
which is then equivalent to 
\begin{align}
 & \left|\left(\mathbf{h}_{k}^{\mathrm{H}}+\mathbf{e}^{\mathrm{H}}\mathbf{G}_{k}\right){\bf f}_{k}\right|^{2}\geq\beta_{k}(2^{R_{k}}-1),\mathcal{E}_{k}^{full},\forall k\in\mathcal{K},\label{eq:single}\\
 & \left\Vert \left(\mathbf{h}_{k}^{\mathrm{H}}+\mathbf{e}^{\mathrm{H}}\mathbf{G}_{k}\right){\bf F}_{-k}\right\Vert _{2}^{2}+\sigma_{k}^{2}\leq\beta_{k},\mathcal{E}_{k}^{full},\forall k\in\mathcal{K}.\label{eq:IN}
\end{align}

The above non-convex semi-infinite inequality constraints can be addressed
in the same way as Scenario 1. In particular, the linear approximation
of the useful signal power in (\ref{eq:single}) is given in the following
lemma.

\begin{lemma}\label{lower-bound-1} Let $\mathbf{f}_{k}^{(n)}$ and
$\mathbf{e}^{(n)}$ be the optimal solutions obtained at iteration
$n$, and by inserting $\mathbf{h}_{k}=\widehat{\mathbf{h}}_{k}+\bigtriangleup\mathbf{h}_{k}$
and $\mathbf{G}_{k}=\widehat{\mathbf{G}}_{k}+\bigtriangleup\mathbf{G}_{k}$
into the useful signal power in (\ref{eq:single}), then the resulting
$|{[({\widehat{\mathbf{h}}_{k}}+{\bigtriangleup\mathbf{h}_{k})^{\mathrm{H}}}+{\mathbf{e}^{\mathrm{H}}}({\widehat{\mathbf{G}}_{k}}+{\bigtriangleup\mathbf{G}_{k}})]}{\bf f}_{k}|^{2}$
is lower bounded linearly at ($\mathbf{f}_{k}^{(n)}$, $\mathbf{e}^{(n)}$)
as follows 
\begin{equation}
\mathbf{x}_{k}^{\mathrm{H}}\widetilde{\mathbf{A}}_{k}\mathbf{x}_{k}+2\mathrm{Re}\left\{ \widetilde{\mathbf{a}}_{k}^{\mathrm{H}}\mathbf{x}_{k}\right\} +\widetilde{a}_{k},\label{eq:lower-bound-full}
\end{equation}
where 
\begin{align*}
\widetilde{\mathbf{A}}_{k} & =\mathbf{D}_{k}+\mathbf{D}_{k}^{\mathrm{H}}-\mathbf{Z}_{k},\\
\mathbf{D}_{k} & =\left[\begin{array}{c}
{\bf f}_{k}^{(n)}\\
{\bf f}_{k}^{(n)}\otimes\mathbf{e}^{\mathrm{(n),*}}
\end{array}\right]\left[\begin{array}{cc}
\mathbf{f}_{k}^{\mathrm{H}} & \mathbf{f}_{k}^{\mathrm{H}}\otimes\mathbf{e}^{\mathrm{T}}\end{array}\right],\\
\mathbf{Z}_{k} & =\left[\begin{array}{c}
{\bf f}_{k}^{(n)}\\
{\bf f}_{k}^{(n)}\otimes\mathbf{e}^{\mathrm{(n),*}}
\end{array}\right]\left[\begin{array}{cc}
\mathbf{f}_{k}^{(n),\mathrm{H}} & \mathbf{f}_{k}^{(n),\mathrm{H}}\otimes\mathbf{e}^{(n),\mathrm{T}}\end{array}\right],\\
\widetilde{\mathbf{a}}_{k} & =\mathbf{d}_{1,k}+\mathbf{d}_{2,k}-\mathbf{z}_{k},\\
\mathbf{d}_{1,k} & =\left[\begin{array}{c}
{\bf f}_{k}\mathbf{f}_{k}^{(n),\mathrm{H}}\left(\mathbf{\widehat{h}}_{k}+\mathbf{\widehat{G}}_{k}^{\mathrm{H}}\mathbf{e}^{(n)}\right)\\
\mathrm{vec}^{*}(\mathbf{e}\left(\mathbf{\widehat{h}}_{k}^{\mathrm{H}}+\mathbf{e}^{(n),\mathrm{H}}\mathbf{\widehat{G}}_{k}\right){\bf f}_{k}^{(n)}\mathbf{f}_{k}^{\mathrm{H}})
\end{array}\right],\\
\mathbf{d}_{2,k} & =\left[\begin{array}{c}
{\bf f}_{k}^{(n)}\mathbf{f}_{k}^{\mathrm{H}}\left(\mathbf{\widehat{h}}_{k}+\mathbf{\widehat{G}}_{k}^{\mathrm{H}}\mathbf{e}\right)\\
\mathrm{vec}^{*}(\mathbf{e}^{(n)}\left(\mathbf{\widehat{h}}_{k}^{\mathrm{H}}+\mathbf{e}^{\mathrm{H}}\mathbf{\widehat{G}}_{k}\right){\bf f}_{k}\mathbf{f}_{k}^{(n),\mathrm{H}})
\end{array}\right],\\
\mathbf{z}_{k} & =\left[\begin{array}{c}
{\bf f}_{k}^{(n)}\mathbf{f}_{k}^{(n),\mathrm{H}}\left(\mathbf{\widehat{h}}_{k}+\mathbf{\widehat{G}}_{k}^{\mathrm{H}}\mathbf{e}^{(n)}\right)\\
\mathrm{vec}^{\mathrm{*}}(\mathbf{e}^{(n)}\left(\mathbf{\widehat{h}}_{k}^{\mathrm{H}}+\mathbf{e}^{(n),\mathrm{H}}\mathbf{\widehat{G}}_{k}\right){\bf f}_{k}^{(n)}\mathbf{f}_{k}^{(n),\mathrm{H}})
\end{array}\right],\\
\widetilde{a}_{k} & =2\mathrm{Re}\left\{ d_{k}\right\} -z_{k},\\
d_{k} & =\left(\mathbf{\widehat{h}}_{k}^{\mathrm{H}}+\mathbf{e}^{(n),\mathrm{H}}\mathbf{\widehat{G}}_{k}\right){\bf f}_{k}^{(n)}\mathbf{f}_{k}^{\mathrm{H}}\left(\mathbf{\widehat{h}}_{k}+\mathbf{\widehat{G}}_{k}^{\mathrm{H}}\mathbf{e}\right),\\
z_{k} & =\left(\mathbf{\widehat{h}}_{k}^{\mathrm{H}}+\mathbf{e}^{(n),\mathrm{H}}\mathbf{\widehat{G}}_{k}\right){\bf f}_{k}^{(n)}\mathbf{f}_{k}^{(n),\mathrm{H}}\left(\mathbf{\widehat{h}}_{k}+\mathbf{\widehat{G}}_{k}^{\mathrm{H}}\mathbf{e}^{(n)}\right),\\
\mathbf{x}_{k} & =\left[\begin{array}{cc}
\bigtriangleup\mathbf{h}_{k}^{\mathrm{H}} & \mathrm{vec}^{\mathrm{H}}(\mathbf{\bigtriangleup\mathbf{G}}_{k}^{*})\end{array}\right]^{\mathrm{H}}.
\end{align*}

\end{lemma}

\textbf{\textit{Proof: }}Please refer to Appendix \ref{subsec:The-proof-of-2}.\hspace{3cm}$\blacksquare$

Based on Lemma \ref{lower-bound-1}, constraints (\ref{eq:single})
are equivalently rewritten as 
\begin{align}
 & \mathbf{x}_{k}^{\mathrm{H}}\widetilde{\mathbf{A}}_{k}\mathbf{x}_{k}+2\mathrm{Re}\left\{ \widetilde{\mathbf{a}}_{k}^{\mathrm{H}}\mathbf{x}_{k}\right\} +\widetilde{a}_{k}\nonumber \\
 & \geq\beta_{k}(2^{R_{k}}-1),\mathcal{E}_{k}^{full},\forall k\in\mathcal{K}.\label{eq:single-full}
\end{align}

Before applying Lemma \ref{S-procedure}, it is beneficial to express
$\mathcal{E}_{k}^{full}$ in terms of the following quadratic expressions
as 
\[
\mathcal{E}_{k}^{full}\triangleq\left\{ \begin{array}{c}
\mathbf{x}_{k}^{\mathrm{H}}\left[\begin{array}{cc}
\mathbf{I}_{N} & \mathbf{0}\\
\mathbf{0} & \mathbf{0}
\end{array}\right]\mathbf{x}_{k}-\xi_{\mathrm{h},k}^{2}\leq0,\\
\mathbf{x}_{k}^{\mathrm{H}}\left[\begin{array}{cc}
\mathbf{0} & \mathbf{0}\\
\mathbf{0} & \mathbf{I}_{MN}
\end{array}\right]\mathbf{x}_{k}-\xi_{\mathrm{g},k}^{2}\leq0.
\end{array}\right.
\]

Therefore, after introducing $\boldsymbol{\varpi}_{\mathrm{h}}=[\varpi_{\mathrm{h},1},...,\varpi_{\mathrm{h},K}]^{\mathrm{T}}\geq0$
and $\boldsymbol{\varpi}_{\mathrm{g}}=[\varpi_{\mathrm{g},1},...,\varpi_{\mathrm{g},K}]^{\mathrm{T}}\geq0$
as slack variables, constraints (\ref{eq:single}) can be transformed
by Lemma \ref{S-procedure} into the following equivalent LMIs as
\begin{align}
 & \left[\begin{array}{cc}
\widetilde{\mathbf{A}}_{k}+\left[\begin{array}{cc}
\varpi_{\mathrm{h},k}\mathbf{I}_{N} & \mathbf{0}\\
\mathbf{0} & \varpi_{\mathrm{g},k}\mathbf{I}_{MN}
\end{array}\right] & \widetilde{\mathbf{a}}_{k}\\
\widetilde{\mathbf{a}}_{k}^{\mathrm{H}} & C_{k}^{full}
\end{array}\right]\succeq\mathbf{0},\forall k\in\mathcal{K},\label{eq:LMI-signal-2}
\end{align}
where $C_{k}^{full}=\widetilde{a}_{k}-\beta_{k}(2^{R_{k}}-1)-\varpi_{\mathrm{h},k}\xi_{\mathrm{h},k}^{2}-\varpi_{\mathrm{g},k}\xi_{\mathrm{g},k}^{2}$.

Next, by inserting $\mathbf{h}_{k}=\widehat{\mathbf{h}}_{k}+\bigtriangleup\mathbf{h}_{k}$
and $\mathbf{G}_{k}=\widehat{\mathbf{G}}_{k}+\bigtriangleup\mathbf{G}_{k}$
into the equivalent matrix inequality of the INs power in (\ref{eq:IN-LMI-1-1}),
we have 
\begin{align}
\mathbf{0} & \preceq\left[\begin{array}{cc}
\beta_{k}-\sigma_{k}^{2} & \widetilde{\mathbf{t}}_{k}^{\mathrm{H}}\\
\widetilde{\mathbf{t}}_{k} & \mathbf{I}
\end{array}\right]\nonumber \\
 & +\left[\begin{array}{cc}
0 & \left(\triangle\mathbf{h}_{k}^{\mathrm{H}}+\mathbf{e}^{\mathrm{H}}\triangle\mathbf{G}_{k}\right){\bf F}_{-k}\\
{\bf F}_{-k}^{\mathrm{H}}\left(\triangle\mathbf{h}_{k}+\triangle\mathbf{G}_{k}^{\mathrm{H}}\mathbf{e}\right) & \mathbf{0}
\end{array}\right]\nonumber \\
 & \preceq\left[\begin{array}{c}
\mathbf{0}\\
{\bf F}_{-k}^{\mathrm{H}}
\end{array}\right]\left[\begin{array}{cc}
\triangle\mathbf{h}_{\mathrm{r},k} & \mathbf{0}\end{array}\right]+\left[\begin{array}{c}
\triangle\mathbf{h}_{\mathrm{r},k}^{\mathrm{H}}\\
\mathbf{0}
\end{array}\right]\left[\begin{array}{cc}
\mathbf{0} & {\bf F}_{-k}\end{array}\right]\nonumber \\
 & +\left[\begin{array}{c}
\mathbf{0}\\
{\bf F}_{-k}^{\mathrm{H}}
\end{array}\right]\triangle\mathbf{G}_{k}^{\mathrm{H}}\left[\begin{array}{cc}
\mathbf{e} & \mathbf{0}\end{array}\right]+\left[\begin{array}{c}
\mathbf{e}^{\mathrm{H}}\\
\mathbf{0}
\end{array}\right]\triangle\mathbf{G}_{k}\left[\begin{array}{cc}
\mathbf{0} & {\bf F}_{-k}\end{array}\right]\nonumber \\
 & +\left[\begin{array}{cc}
\beta_{k}-\sigma_{k}^{2} & \widetilde{\mathbf{t}}_{k}^{\mathrm{H}}\\
\widetilde{\mathbf{t}}_{k} & \mathbf{I}
\end{array}\right],\label{eq:IN-LMI-FULL-2}
\end{align}
where $\widetilde{\mathbf{t}}_{k}=((\widehat{\mathbf{h}}_{k}^{\mathrm{H}}+\mathbf{e}^{\mathrm{H}}\widehat{\mathbf{G}}_{k}){\bf F}_{-k})^{\mathrm{H}}$.

Applying Lemma \ref{sign-definiteness} and defining slack variables
$\boldsymbol{\mu}_{\mathrm{g}}=[\mu_{\mathrm{g},1},...,\mu_{\mathrm{g},K}]^{\mathrm{T}}\geq0$
and $\boldsymbol{\mu}_{\mathrm{h}}=[\mu_{\mathrm{h},1},...,\mu_{\mathrm{h},K}]^{\mathrm{T}}\geq0$,
the equivalent LMIs of the worst-case INs power constraints (\ref{eq:IN})
are given as 
\begin{align}
 & \left[\begin{array}{cccc}
Temp_{k} & \widetilde{\mathbf{t}}_{k}^{\mathrm{H}} & \mathbf{0}_{1\times N} & \mathbf{0}_{1\times N}\\
\widetilde{\mathbf{t}}_{k} & \mathbf{I}_{(K-1)} & \xi_{\mathrm{g},k}{\bf F}_{-k}^{\mathrm{H}} & \xi_{\mathrm{h},k}{\bf F}_{-k}^{\mathrm{H}}\\
\mathbf{0}_{N\times1} & \xi_{\mathrm{g},k}{\bf F}_{-k} & \mu_{\mathrm{g},k}\mathbf{I}_{N} & \mathbf{0}_{N\times N}\\
\mathbf{0}_{N\times1} & \xi_{\mathrm{h},k}{\bf F}_{-k} & \mathbf{0}_{N\times N} & \mu_{\mathrm{h},k}\mathbf{I}_{N}
\end{array}\right]\succeq\mathbf{0},\forall k\in\mathcal{K},\label{eq:LMI-IN}
\end{align}
where $Temp_{k}=\beta_{k}-\sigma_{k}^{2}-\mu_{\mathrm{g},k}M-\mu_{\mathrm{h},k}$.

With (\ref{eq:LMI-signal-2}) and (\ref{eq:LMI-IN}), the worst-case
robust beamforming design problem under full channel uncertainty can
be formulated as \begin{subequations}\label{Pro:min-power-worst-full-1}
\begin{align}
\min\limits _{{\scriptstyle {\mathbf{F},\mathbf{e},\boldsymbol{\beta},\boldsymbol{\varpi}_{\mathrm{g}},\hfill\atop {\scriptstyle \boldsymbol{\varpi}_{\mathrm{h}},\boldsymbol{\mu}_{\mathrm{g}}\boldsymbol{\mu}_{\mathrm{h}}\hfill}}}} & \;\;||\mathbf{F}||_{F}^{2}\label{eq:obj-2-1}\\
{\rm s.t.} & \thinspace\thinspace\thinspace(\ref{eq:LMI-signal-2}),(\ref{eq:LMI-IN}),(\ref{eq:min-power-cons2}),\\
 & \thinspace\thinspace\thinspace\boldsymbol{\varpi}_{\mathrm{g}}\geq0,\boldsymbol{\varpi}_{\mathrm{h}}\geq0,\boldsymbol{\mu}_{\mathrm{g}}\geq0,\boldsymbol{\mu}_{\mathrm{h}}\geq0.\label{eq:omiga-1}
\end{align}
\end{subequations} Problem (\ref{Pro:min-power-worst-full-1}) is
again non-convex and has coupled variables, which can be solved similarly
to Problem (\ref{Pro:min-power-worst-partial-2}) and thus omitted
for simplicity.

\section{Outage constrained robust beamforming design}

In general, the channel estimation error follows the Gaussian distribution
\cite{shuguang-IRS}. Hence, it is unbounded. The above bounded channel
model may not be able to characterize the practical channel error
model. As a result, in this section, we consider the statistical CSI
error model. Specifically, by defining the maximum data rate outage
probabilities $\rho_{1},...,\rho_{K}\in(0,1]$, the transmit power
minimization problem is formulated as \begin{subequations}\label{Pro:min-power}
\begin{align}
\mathop{\min}\limits _{\mathbf{F},\mathbf{e}} & \;\;||\mathbf{F}||_{F}^{2}\label{eq:min-power-obj}\\
\textrm{s.t.} & \thinspace\thinspace\thinspace\mathrm{Pr}\{\mathcal{R}_{k}\left(\mathbf{F},\mathbf{e}\right)\geq R_{k}\}\geq1-\rho_{k},\forall k\in\mathcal{K}\thinspace\label{eq:min-power-cons1}\\
 & \thinspace\thinspace\thinspace|e_{m}|^{2}=1,1\leq m\leq M.
\end{align}
\end{subequations}The rate outage constraints (\ref{eq:min-power-cons1})
guarantee that the probability of each user that can successfully
decode its message at a data rate of $R_{k}$ is no less than $1-\rho_{k}$.

The outage constrained robust beamforming design problem in (\ref{Pro:min-power})
is computationally intractable due to the fact that the rate outage
probability constraints (\ref{eq:min-power-cons1}) have no simple
closed-form expressions \cite{W-K-MA2014}. In order to solve Problem
(\ref{Pro:min-power}), a safe approximation based on Bernstein-type
inequality is given in the following lemma.

\begin{lemma}\label{Bernstein-Type Inequalities} (Bernstein-Type
Inequality: Lemma 1 in \cite{W-K-MA2014}) Assume $f(\mathbf{x})=\mathbf{x}^{\mathrm{H}}\mathbf{U}\mathbf{x}+2\mathrm{Re}\{\mathbf{u}^{\mathrm{H}}\mathbf{x}\}+u$,
where $\mathbf{U}\in\mathbb{H}^{n\times n}$, $\mathbf{u}\in\mathbb{C}^{n\times1}$,
$u\in\mathbb{R}$ and $\mathbf{x}\in\mathbb{C}^{n\times1}\sim\mathcal{CN}(\mathbf{0},\mathbf{I})$.
Then for any $\rho\in[0,1]$, the following approximation holds: \begin{subequations}\label{eq:Outage-5}
\begin{align}
 & \mathrm{Pr}\{\mathbf{x}^{\mathrm{H}}\mathbf{U}\mathbf{x}+2\mathrm{Re}\{\mathbf{u}^{\mathrm{H}}\mathbf{x}\}+u\geq0\}\geq1-\rho\\
\Rightarrow & \mathrm{Tr}\left\{ \mathbf{U}\right\} -\sqrt{2\ln(1/\rho)}x+\ln(\rho)\lambda_{\max}^{+}(-\mathbf{U})+u\geq0\\
\Rightarrow & \left\{ \begin{array}{c}
\mathrm{Tr}\left\{ \mathbf{U}\right\} -\sqrt{2\ln(1/\rho)}x+\ln(\rho)y+u\geq0\\
\sqrt{||\mathbf{U}||_{F}^{2}+2||\mathbf{u}||^{2}}\leq x\\
y\mathbf{I}+\mathbf{U}\succeq\mathbf{0},y\geq0,
\end{array}\right.
\end{align}
where $\lambda_{\max}^{+}(-\mathbf{U})=\max(\lambda_{\max}(-\mathbf{U}),0)$.
$x$ and $y$ are slack variables. \end{subequations} \end{lemma}
Please refer to \cite{W-K-MA2014} for the proof of Lemma \ref{Bernstein-Type Inequalities}.

In the following subsections, we first design the relatively simple
robust beamforming under the partial channel uncertainty, and then
extend it to the full channel uncertainty case.

\subsection{Scenario 1: Partial Channel Uncertainty}

Before the derivations, the rate outage probability of user $k$ in
(\ref{eq:min-power-cons1}) is rewritten as 
\begin{align}
 & \mathrm{Pr}\left\{ \log_{2}\left(1+\frac{\left|\left(\mathbf{h}_{k}^{\mathrm{H}}+\mathbf{e}^{\mathrm{H}}\mathbf{G}_{k}\right){\bf f}_{k}\right|^{2}}{\left\Vert \left(\mathbf{h}_{k}^{\mathrm{H}}+\mathbf{e}^{\mathrm{H}}\mathbf{G}_{k}\right){\bf F}_{-k}\right\Vert _{2}^{2}+\sigma_{k}^{2}}\right)\geq R_{k}\right\} \nonumber \\
 & =\mathrm{Pr}\left\{ \left(\mathbf{h}_{k}^{\mathrm{H}}+\mathbf{e}^{\mathrm{H}}\mathbf{G}_{k}\right)\boldsymbol{\Phi}_{k}\left(\mathbf{h}_{k}+\mathbf{G}_{k}^{\mathrm{H}}\mathbf{e}\right)-\sigma_{k}^{2}\geq0\right\} ,\label{eq:Outage-1}
\end{align}
where $\boldsymbol{\Phi}_{k}={\bf f}_{k}\mathbf{f}_{k}^{\mathrm{H}}/(2^{R_{k}}-1)-{\bf F}_{-k}\mathbf{F}_{-k}^{\mathrm{H}}$.

For the convenience of derivations, we assume that $\boldsymbol{\Sigma}_{\mathrm{g},k}=\varepsilon_{\mathrm{g},k}^{2}\mathbf{I}$,
then the RCSIT error in (\ref{Pro:statistic-error}) can be rewritten
as $\mathrm{vec}(\bigtriangleup\mathbf{G}_{k})=\varepsilon_{\mathrm{g},k}\mathbf{i}_{\mathrm{g},k}$
where $\mathbf{i}_{\mathrm{g},k}\sim\mathcal{CN}(\mathbf{0},\mathbf{I})$.
Defining $\mathbf{E}=\mathbf{e}\mathbf{e}^{\mathrm{H}}$, the rate
outage probability (\ref{eq:Outage-1}) is reformulated in (\ref{eq:Outage-2})
\begin{figure*}
\begin{align}
 & \mathrm{Pr}\left\{ \left(\mathbf{h}_{k}^{\mathrm{H}}+\mathbf{e}^{\mathrm{H}}(\mathbf{\widehat{G}}_{k}+\bigtriangleup\mathbf{G}_{k})\right)\boldsymbol{\Phi}_{k}\left(\mathbf{h}_{k}+(\mathbf{\widehat{G}}_{k}+\bigtriangleup\mathbf{G}_{k})^{\mathrm{H}}\mathbf{e}\right)-\sigma_{k}^{2}\geq0\right\} \nonumber \\
 & =\mathrm{Pr}\Bigl\{\mathrm{vec}^{\mathrm{H}}(\bigtriangleup\mathbf{G}_{k})(\boldsymbol{\Phi}_{k}^{\mathrm{T}}\otimes\mathbf{E})\mathrm{vec}(\bigtriangleup\mathbf{G}_{k})+2\mathrm{Re}\{\mathrm{vec}^{\mathrm{H}}((\mathbf{e}\mathbf{h}_{k}^{\mathrm{H}}+\mathbf{E}\mathbf{\widehat{G}}_{k})\boldsymbol{\Phi}_{k})\mathrm{vec}(\mathbf{\bigtriangleup\mathbf{G}}_{k})\}\nonumber \\
 & \thinspace\thinspace\thinspace\thinspace\thinspace\thinspace\thinspace+(\mathbf{h}_{k}^{\mathrm{H}}+\mathbf{e}^{\mathrm{H}}\mathbf{\widehat{G}}_{k})\boldsymbol{\Phi}_{k}(\mathbf{h}_{k}+\mathbf{\widehat{G}}_{k}^{\mathrm{H}}\mathbf{e})-\sigma_{k}^{2}\geq0\Bigr\}\nonumber \\
 & =\mathrm{Pr}\left\{ \varepsilon_{\mathrm{g},k}^{2}\mathbf{i}_{\mathrm{g},k}^{\mathrm{H}}(\boldsymbol{\Phi}_{k}^{\mathrm{T}}\otimes\mathbf{E})\mathbf{i}_{\mathrm{g},k}+2\mathrm{Re}\{\varepsilon_{\mathrm{g},k}\mathrm{vec}^{\mathrm{H}}((\mathbf{e}\mathbf{h}_{k}^{\mathrm{H}}+\mathbf{E}\mathbf{\widehat{G}}_{k})\boldsymbol{\Phi}_{k})\mathbf{i}_{\mathrm{g},k}\}+(\mathbf{h}_{k}^{\mathrm{H}}+\mathbf{e}^{\mathrm{H}}\mathbf{\widehat{G}}_{k})\boldsymbol{\Phi}_{k}(\mathbf{h}_{k}+\mathbf{\widehat{G}}_{k}^{\mathrm{H}}\mathbf{e})-\sigma_{k}^{2}\geq0\right\} .\label{eq:Outage-2}
\end{align}
\hrule 
\end{figure*}

at the top of the next page. Therefore, the rate outage constraints
(\ref{eq:min-power-cons1}) are given as 
\begin{align}
 & \mathrm{Pr}\left\{ \mathbf{i}_{\mathrm{g},k}^{\mathrm{H}}\mathbf{U}_{k}\mathbf{i}_{\mathrm{g},k}+2\mathrm{Re}\{\mathbf{u}_{k}^{\mathrm{H}}\mathbf{i}_{\mathrm{g},k}\}+u_{k}\geq0\right\} \nonumber \\
 & \thinspace\thinspace\thinspace\thinspace\thinspace\thinspace\thinspace\thinspace\geq1-\rho_{k},\forall k\in\mathcal{K},\label{eq:Outage-3}
\end{align}
where \begin{subequations} 
\begin{align}
\mathbf{U}_{k} & =\varepsilon_{\mathrm{g},k}^{2}(\boldsymbol{\Phi}_{k}^{\mathrm{T}}\otimes\mathbf{E}),\\
\mathbf{u}_{k} & =\varepsilon_{\mathrm{g},k}\mathrm{vec}((\mathbf{e}\mathbf{h}_{k}^{\mathrm{H}}+\mathbf{E}\mathbf{\widehat{G}}_{k})\boldsymbol{\Phi}_{k}^{\mathrm{H}}),\\
u_{k} & =(\mathbf{h}_{k}^{\mathrm{H}}+\mathbf{e}^{\mathrm{H}}\mathbf{\widehat{G}}_{k})\boldsymbol{\Phi}_{k}(\mathbf{h}_{k}+\mathbf{\widehat{G}}_{k}^{\mathrm{H}}\mathbf{e})-\sigma_{k}^{2}.\label{eq:c}
\end{align}
\end{subequations}

Applying Lemma \ref{Bernstein-Type Inequalities} and introducing
auxiliary variables $\mathbf{x}=[x_{1},...,x_{K}]^{\mathrm{T}}$ and
$\mathbf{y}=[y_{1},...,y_{K}]^{\mathrm{T}}$, rate outage constraint
of user $k$ in (\ref{eq:Outage-3}) is transformed into the deterministic
form as \begin{subequations}\label{eq:Outage-6} 
\begin{align}
 & \mathrm{Tr}\left\{ \mathbf{U}_{k}\right\} -\sqrt{2\ln(1/\rho_{k})}x_{k}+\ln(\rho_{k})y_{k}+u_{k}\geq0,\label{eq:BTI-1}\\
 & \sqrt{||\mathbf{U}_{k}||_{F}^{2}+2||\mathbf{u}_{k}|||^{2}}\leq x_{k},\label{eq:BTI-2}\\
 & y_{k}\mathbf{I}+\mathbf{U}_{k}\succeq\mathbf{0},y_{k}\geq0.\label{eq:BTI-3}
\end{align}
\end{subequations}

(\ref{eq:Outage-6}) can be further simplified by some mathematical
transformations as follows \begin{subequations}\label{eq:Outage-7}
\begin{align}
 & \mathrm{Tr}\left\{ \mathbf{U}_{k}\right\} =\varepsilon_{\mathrm{g},k}^{2}\mathrm{Tr}\left\{ \boldsymbol{\Phi}_{k}^{\mathrm{T}}\otimes\mathbf{E}\right\} =\varepsilon_{\mathrm{g},k}^{2}\mathrm{Tr}\left\{ \boldsymbol{\Phi}_{k}\right\} \mathrm{Tr}\left\{ \mathbf{E}\right\} \nonumber \\
 & \qquad\quad\;\;=\varepsilon_{\mathrm{g},k}^{2}M\mathrm{Tr}\left\{ \boldsymbol{\Phi}_{k}\right\} ,\label{eq:BTI-1-1}\\
 & ||\mathbf{U}_{k}||_{F}^{2}=\varepsilon_{\mathrm{g},k}^{4}||(\boldsymbol{\Phi}_{k}^{\mathrm{T}}\otimes\mathbf{E})||_{F}^{2}=\varepsilon_{\mathrm{g},k}^{4}||\boldsymbol{\Phi}_{k}||_{F}^{2}||\mathbf{E}||_{F}^{2}\nonumber \\
 & \qquad\;\;\;=\varepsilon_{\mathrm{g},k}^{4}M^{2}||\boldsymbol{\Phi}_{k}||_{F}^{2},\label{eq:BTI-2-1}\\
 & ||\mathbf{u}_{k}||^{2}=\varepsilon_{\mathrm{g},k}^{2}||\mathrm{vec}((\mathbf{e}\mathbf{h}_{k}^{\mathrm{H}}+\mathbf{E}\mathbf{\widehat{G}}_{k})\boldsymbol{\Phi}_{k}^{\mathrm{H}})||^{2}\nonumber \\
 & \qquad\;\thinspace=\varepsilon_{\mathrm{g},k}^{2}M||\left(\mathbf{h}_{k}^{\mathrm{H}}+\mathbf{e}^{\mathrm{H}}\mathbf{\widehat{G}}_{k}\right)\boldsymbol{\Phi}_{k}||_{2}^{2},\label{eq:BTI-3-1}\\
 & \lambda(\mathbf{U}_{k})=\lambda(\varepsilon_{\mathrm{g},k}^{2}(\boldsymbol{\Phi}_{k}^{\mathrm{T}}\otimes\mathbf{E}))=\varepsilon_{\mathrm{g},k}^{2}\lambda(\boldsymbol{\Phi}_{k}^{\mathrm{T}}\otimes\mathbf{E})\nonumber \\
 & \qquad\;\;\thinspace=\varepsilon_{\mathrm{g},k}^{2}\lambda(\boldsymbol{\Phi}_{k})\lambda(\mathbf{E})=\varepsilon_{\mathrm{g},k}^{2}M\lambda(\boldsymbol{\Phi}_{k}).\label{eq:BTI-4-1}
\end{align}
\end{subequations}Operation $\lambda(\mathbf{X})$ means the eigenvalues
of $\mathbf{X}$. (\ref{eq:BTI-1-1}) and (\ref{eq:BTI-2-1}) are
from {[}P76 in \cite{Xinda2017}{]}, (\ref{eq:BTI-4-1}) is from {[}P421
in \cite{Xinda2017}{]}.

Therefore, according to Lemma \ref{Bernstein-Type Inequalities} and
equation (\ref{eq:Outage-7}), the approximation problem of Problem
(\ref{Pro:min-power}) can be given as

\begin{subequations}\label{Pro:min-power-1} 
\begin{align}
\mathop{\min}\limits _{\mathbf{F},\mathbf{e},\mathbf{x},\mathbf{y}} & \;||\mathbf{F}||_{F}^{2}\\
\textrm{s.t.} & \;\;\varepsilon_{\mathrm{g},k}^{2}M\mathrm{Tr}\left\{ \boldsymbol{\Phi}_{k}\right\} -\sqrt{2\ln(1/\rho_{k})}x_{k}-\ln(1/\rho_{k})y_{k}\nonumber \\
 & \thinspace\thinspace+u_{k}\geq0,\forall k\in\mathcal{K}\label{eq:cons-1-1-1}\\
 & \;\;\left\Vert \begin{array}{c}
\varepsilon_{\mathrm{g},k}^{2}M\mathrm{vec}(\boldsymbol{\Phi}_{k})\\
\sqrt{2M}\varepsilon_{\mathrm{g},k}\boldsymbol{\Phi}_{k}\left(\mathbf{h}_{k}+\mathbf{\widehat{G}}_{k}^{\mathrm{H}}\mathbf{e}\right)
\end{array}\right\Vert \leq x_{k},\forall k\in\mathcal{K}\label{eq:cons-2-1-1}\\
 & \;\;y_{k}\mathbf{I}+\varepsilon_{\mathrm{g},k}^{2}M\boldsymbol{\Phi}_{k}\succeq\mathbf{0},y_{k}\geq0,\forall k\in\mathcal{K}\label{eq:cons-3-1-1}\\
 & \;\;|e_{m}|^{2}=1,\forall m\in\mathcal{M}.\label{eq:unit-cons}
\end{align}
\end{subequations}

Problem (\ref{Pro:min-power-1}) is still difficult to solve because
constraints (\ref{eq:cons-2-1-1}) are non-convex and have coupled
variables $\mathbf{F}$ and $\mathbf{e}$. We use AO method to update
$\mathbf{F}$ and $\mathbf{e}$ in an iterative manner. More specifically,
by first fixing $\mathbf{e}$, the non-convex problem in $\mathbf{F}$
at hand is relaxed by employing the SDR technique \cite{luo2010SDR}
and solved by CVX. $\mathbf{F}$ is then fixed and the resulting non-convex
problem of $\mathbf{e}$ is also handled under the penalty CCP method.

For fixed $\mathbf{e}$, let $\boldsymbol{\Phi}_{k}=\boldsymbol{\Gamma}_{k}/(2^{R_{k}}-1)-\sum_{i=1,i\neq k}^{K}\boldsymbol{\Gamma}_{i}$
where $\boldsymbol{\Gamma}_{k}=\mathbf{f}_{k}\mathbf{f}_{k}^{\mathrm{H}}$,
Problem (\ref{Pro:min-power-1}) corresponding to $\mathbf{F}$ is
rewritten as \begin{subequations}\label{Pro:min-power-f} 
\begin{align}
\mathop{\min}\limits _{\boldsymbol{\Gamma},\mathbf{x},\mathbf{y}} & \;\sum_{k=1}^{K}\mathrm{Tr}\left\{ \boldsymbol{\Gamma}_{k}\right\} \\
\textrm{s.t.} & \;\;\varepsilon_{\mathrm{g},k}^{2}M\mathrm{Tr}\left\{ \boldsymbol{\Phi}_{k}\right\} -\sqrt{2\ln(1/\rho_{k})}x_{k}-\ln(1/\rho_{k})y_{k}\nonumber \\
 & \thinspace\thinspace\thinspace+u_{k}\geq0,\forall k\in\mathcal{K}\label{eq:cons-1-1}\\
 & \;\;\left\Vert \begin{array}{c}
\varepsilon_{\mathrm{g},k}^{2}M\mathrm{vec}(\boldsymbol{\Phi}_{k})\\
\sqrt{2M}\varepsilon_{\mathrm{g},k}\boldsymbol{\Phi}_{k}\left(\mathbf{h}_{k}+\mathbf{\widehat{G}}_{k}^{\mathrm{H}}\mathbf{e}\right)
\end{array}\right\Vert \leq x_{k},\forall k\in\mathcal{K}\label{eq:cons-2-1}\\
 & \;\;y_{k}\mathbf{I}+\varepsilon_{\mathrm{g},k}^{2}M\boldsymbol{\Phi}_{k}\succeq\mathbf{0},y_{k}\geq0,\forall k\in\mathcal{K}\label{eq:cons-3-1}\\
 & \;\;\boldsymbol{\Gamma}_{k}\succeq\mathbf{0},\forall k\in\mathcal{K}\label{eq:sdf-1}\\
 & \;\;\mathrm{rank}(\boldsymbol{\Gamma}_{k})=1,\forall k\in\mathcal{K},
\end{align}
\end{subequations} where $\boldsymbol{\Gamma}=[\boldsymbol{\Gamma}_{1},...,\boldsymbol{\Gamma}_{K}]$.
Problem (\ref{Pro:min-power-f}) can be solved by adopting the SDR
technique, i.e., removing $\mathrm{rank}(\boldsymbol{\Gamma}_{k})=1,\forall k\in\mathcal{K}$
from the problem formulation, the resulting convex SDP problem is
then efficiently solved by the CVX tools. The following theorem reveals
the tightness of the SDR.

\begin{theorem}\label{Theorem-1} If the relaxed version of Problem
(\ref{Pro:min-power-f}) is feasible, then there always exists a feasible
solution, denoted as $\boldsymbol{\Gamma}^{\star}=[\boldsymbol{\Gamma}_{1}^{\star},...,\boldsymbol{\Gamma}_{K}^{\star}]$,
satisfying $\mathrm{rank}(\boldsymbol{\Gamma}_{k}^{\star})=1,\forall k\in\mathcal{K}$.

\end{theorem}

\textbf{\textit{Proof: }}Please refer to Appendix \ref{subsec:The-proof-of-rank-1}.\hspace{3cm}$\blacksquare$

\textit{Remark 1: }Numerical results show that, the optimal $\boldsymbol{\Gamma}_{k}^{\star}$
is usually of rank one before we construct the rank-1 solution mentioned
in Appendix \ref{subsec:The-proof-of-rank-1}. The optimal $\mathbf{f}_{k}$
can be obtained from $\boldsymbol{\Gamma}_{k}^{\star}$ by using eigenvalue
decomposition.

We now consider the subproblem of $\mathbf{e}$ with fixed $\mathbf{F}$.
With the same purpose of (\ref{eq:SINR-slac}), we introduce slack
variables $\boldsymbol{\alpha}=[\alpha_{1},...,\alpha_{K}]^{\mathrm{T}}$
to the rate outage probability in (\ref{eq:Outage-1}) and have 
\begin{align}
 & \mathrm{Pr}\left\{ \left(\mathbf{h}_{k}^{\mathrm{H}}+\mathbf{e}^{\mathrm{H}}\mathbf{G}_{k}\right)\boldsymbol{\Phi}_{k}\left(\mathbf{h}_{k}+\mathbf{G}_{k}^{\mathrm{H}}\mathbf{e}\right)-\sigma_{k}^{2}-\alpha_{k}\geq0\right\} .\label{eq:Outage-1-1}
\end{align}
Then, (\ref{eq:c}) is also modified as follows 
\begin{equation}
u_{k}^{\mathbf{e}}=(\mathbf{h}_{k}^{\mathrm{H}}+\mathbf{e}^{\mathrm{H}}\mathbf{\widehat{G}}_{k})\boldsymbol{\Phi}_{k}(\mathbf{h}_{k}+\mathbf{\widehat{G}}_{k}^{\mathrm{H}}\mathbf{e})-\sigma_{k}^{2}-\alpha_{k}.\label{eq:c-e}
\end{equation}
We note that (\ref{eq:c-e}) is non-concave in $\mathbf{e}$ due to
the fact that $\mathbf{e}^{\mathrm{H}}\mathbf{\widehat{G}}_{k}\mathbf{f}_{k}\mathbf{f}_{k}^{\mathrm{H}}\mathbf{\widehat{G}}_{k}^{\mathrm{H}}\mathbf{e}/(2^{R_{k}}-1)$
in $\mathbf{\widehat{G}}_{k}\boldsymbol{\Phi}_{k}\mathbf{\widehat{G}}_{k}^{\mathrm{H}}$
is convex. By using the first-order Taylor inequality given in Appendix
\ref{subsec:The-proof-of-1}, $\mathbf{e}^{\mathrm{H}}\mathbf{\widehat{G}}_{k}\mathbf{f}_{k}\mathbf{f}_{k}^{\mathrm{H}}\mathbf{\widehat{G}}_{k}^{\mathrm{H}}\mathbf{e}/(2^{R_{k}}-1)$
can be lower bounded linearly by $u_{\mathrm{linear},k}^{\mathbf{e}}=(2\mathrm{Re}\{\mathbf{e}^{(n),\mathrm{H}}\mathbf{\widehat{G}}_{k}\mathbf{f}_{k}\mathbf{f}_{k}^{\mathrm{H}}\mathbf{\widehat{G}}_{k}^{\mathrm{H}}\mathbf{e}\}-\mathbf{e}^{(n),\mathrm{H}}\mathbf{\widehat{G}}_{k}\mathbf{f}_{k}\mathbf{f}_{k}^{\mathrm{H}}\mathbf{\widehat{G}}_{k}^{\mathrm{H}}\mathbf{e}^{(n)})/(2^{R_{k}}-1)$.
We then construct an equivalent concave version of (\ref{eq:c-e}),
which is given as 
\begin{align}
u_{k}^{\mathbf{e}} & =u_{\mathrm{linear},k}^{\mathbf{e}}-\mathbf{e}^{\mathrm{H}}\widehat{\mathbf{G}}_{k}{\bf F}_{-k}\mathbf{F}_{-k}^{\mathrm{H}}\widehat{\mathbf{G}}_{k}^{\mathrm{H}}\mathbf{e}+2\mathrm{Re}\{\mathbf{e}^{\mathrm{H}}\widehat{\mathbf{G}}_{k}\boldsymbol{\Phi}_{k}\mathbf{h}_{k}\}\nonumber \\
 & \thinspace\thinspace\thinspace\thinspace\thinspace\thinspace+\mathbf{h}_{k}^{\mathrm{H}}\boldsymbol{\Phi}_{k}\mathbf{h}_{k}-\sigma_{k}^{2}-\alpha_{k}+M\mathrm{constE}_{k}.\label{eq:concave-c}
\end{align}

In addition, constraints (\ref{eq:cons-3-1-1}) are independent of
$\mathbf{e}$ and transformed from $\lambda_{\max}^{+}(-\mathbf{U})$
in Lemma \ref{Bernstein-Type Inequalities}, we then can have $y_{k}=\max(\lambda_{\max}(-\varepsilon_{\mathrm{g},k}^{2}M\boldsymbol{\Phi}_{k}),0),\forall k\in\mathcal{K}$.
With $\boldsymbol{\alpha}$ and (\ref{eq:concave-c}), the subproblem
of (\ref{Pro:min-power-1}) corresponding to $\mathbf{e}$ is formulated
as \begin{subequations}\label{Pro:min-power-e-1} 
\begin{align}
\mathop{\max}\limits _{\mathbf{e},\boldsymbol{\alpha},\mathbf{x},\mathbf{y}} & \;\sum_{k=1}^{K}\alpha_{k}\\
\textrm{s.t.} & \;\;\varepsilon_{\mathrm{g},k}^{2}M\mathrm{Tr}\left\{ \boldsymbol{\Phi}_{k}\right\} -\sqrt{2\ln(1/\rho_{k})}x_{k}\nonumber \\
 & \thinspace\thinspace\thinspace-\ln(1/\rho_{k})y_{k}+u_{k}^{\mathbf{e}}\geq0,\forall k\in\mathcal{K}\\
 & \;\;\left\Vert \begin{array}{c}
\varepsilon_{\mathrm{g},k}^{2}M||\boldsymbol{\Phi}_{k}||_{F}\\
\sqrt{2M}\varepsilon_{\mathrm{g},k}\boldsymbol{\Phi}_{k}\left(\mathbf{h}_{k}+\mathbf{\widehat{G}}_{k}^{\mathrm{H}}\mathbf{e}\right)
\end{array}\right\Vert \leq x_{k},\forall k\in\mathcal{K}\\
 & \;\;\boldsymbol{\alpha}\geq0\\
 & \;\;|e_{m}|^{2}=1,\forall m\in\mathcal{M}.\label{eq:sdf}
\end{align}
\end{subequations}The non-convex constraints (\ref{eq:sdf}) in Problem
(\ref{Pro:min-power-e-1}) is solved by using the same techniques
as those used for solving Problem (\ref{Pro:min-power-4}), then the
resulting approximation problem for Problem (\ref{Pro:min-power-e-1})
can be solved by using Algorithm \ref{Algorithm-analog}.

\subsection{Scenario 2: Full Channel Uncertainty}

In this subsection, we extend the outage constrained robust beamforming
design from the partial channel uncertainty to the case where all
the channels are imperfect at the BS. By considering the full statistical
CSI error in (\ref{Pro:statistic-error}), (\ref{eq:Outage-1}) is
then formulated as

\begin{align}
 & \mathrm{Pr}\Bigl\{\left(\mathbf{\widehat{h}}_{k}^{\mathrm{H}}+\mathbf{e}^{\mathrm{H}}\mathbf{\widehat{G}}_{k}\right)\boldsymbol{\Phi}_{k}\left(\mathbf{\widehat{h}}_{k}+\mathbf{\widehat{G}}_{k}^{\mathrm{H}}\mathbf{e}\right)\nonumber \\
 & +2\mathrm{Re}\left\{ \left(\mathbf{\widehat{h}}_{k}^{\mathrm{H}}+\mathbf{e}^{\mathrm{H}}\mathbf{\widehat{G}}_{k}\right)\boldsymbol{\Phi}_{k}\left(\bigtriangleup\mathbf{h}_{k}+\mathbf{\bigtriangleup\mathbf{G}}_{k}^{\mathrm{H}}\mathbf{e}\right)\right\} -\sigma_{k}^{2}\nonumber \\
 & +\left(\mathbf{\bigtriangleup\mathbf{h}}_{k}^{\mathrm{H}}+\mathbf{e}^{\mathrm{H}}\bigtriangleup\mathbf{G}_{k}\right)\boldsymbol{\Phi}_{k}\left(\bigtriangleup\mathbf{h}_{k}+\mathbf{\bigtriangleup\mathbf{G}}_{k}^{\mathrm{H}}\mathbf{e}\right)\geq0\Bigr\}.\label{eq:Outage-D-1}
\end{align}
Assuming that $\boldsymbol{\Sigma}_{\mathrm{h},k}=\varepsilon_{\mathrm{h},k}^{2}\mathbf{I}$,
then the DCSIT can be expressed as $\bigtriangleup\mathbf{h}_{k}=\varepsilon_{\mathrm{h},k}\mathbf{i}_{\mathrm{h},k}$
where $\mathbf{i}_{\mathrm{h},k}\sim\mathcal{CN}(\mathbf{0},\mathbf{I})$.
The second term inside (\ref{eq:Outage-D-1}) is rewritten as 
\begin{align*}
 & 2\mathrm{Re}\Bigl\{(\mathbf{\widehat{h}}_{k}^{\mathrm{H}}+\mathbf{e}^{\mathrm{H}}\mathbf{\widehat{G}}_{k})\boldsymbol{\Phi}_{k}\bigtriangleup\mathbf{h}_{k}\\
 & \thinspace\thinspace\thinspace\thinspace\thinspace\thinspace+\mathrm{vec}^{\mathrm{T}}(\mathbf{e}(\mathbf{\widehat{h}}_{k}^{\mathrm{H}}+\mathbf{e}^{\mathrm{H}}\mathbf{\widehat{G}}_{k})\boldsymbol{\Phi}_{k})\mathrm{vec}(\mathbf{\bigtriangleup\mathbf{G}}_{k}^{*})\Bigr\}\\
 & =2\mathrm{Re}\Bigl\{\varepsilon_{\mathrm{h},k}(\mathbf{\widehat{h}}_{k}^{\mathrm{H}}+\mathbf{e}^{\mathrm{H}}\mathbf{\widehat{G}}_{k})\boldsymbol{\Phi}_{k}\mathbf{i}_{\mathrm{h},k}\\
 & \thinspace\thinspace\thinspace\thinspace\thinspace\thinspace+\varepsilon_{\mathrm{g},k}\mathrm{vec}^{\mathrm{T}}(\mathbf{e}(\mathbf{\widehat{h}}_{k}^{\mathrm{H}}+\mathbf{e}^{\mathrm{H}}\mathbf{\widehat{G}}_{k})\boldsymbol{\Phi}_{k})\mathbf{i}_{\mathrm{g},k}^{*}\Bigr\}\\
 & =2\mathrm{Re}\left\{ \widetilde{\mathbf{u}}_{k}^{\mathrm{H}}\widetilde{\mathbf{i}}_{k}\right\} ,
\end{align*}
where $\widetilde{\mathbf{i}}_{k}=\left[\begin{array}{cc}
\mathbf{i}_{\mathrm{h},k}^{\mathrm{H}} & \mathbf{i}_{\mathrm{g},k}^{\mathrm{T}}\end{array}\right]^{\mathrm{H}}$ and 
\[
\widetilde{\mathbf{u}}_{k}=\left[\begin{array}{c}
\varepsilon_{\mathrm{h},k}\boldsymbol{\Phi}_{k}(\mathbf{\widehat{h}}_{k}+\mathbf{\widehat{G}}_{k}^{\mathrm{H}}\mathbf{e})\\
\varepsilon_{\mathrm{g},k}\mathrm{vec}^{*}(\mathbf{e}(\mathbf{\widehat{h}}_{k}^{\mathrm{H}}+\mathbf{e}^{\mathrm{H}}\mathbf{\widehat{G}}_{k})\boldsymbol{\Phi}_{k})
\end{array}\right].
\]
The fourth term on the left hand side of (\ref{eq:Outage-D-1}) is
rewritten as 
\begin{align*}
 & \mathbf{\bigtriangleup\mathbf{h}}_{k}^{\mathrm{H}}\boldsymbol{\Phi}_{k}\bigtriangleup\mathbf{h}_{k}+2\mathrm{Re}\left\{ \mathbf{e}^{\mathrm{H}}\bigtriangleup\mathbf{G}_{k}\boldsymbol{\Phi}_{k}\bigtriangleup\mathbf{h}_{k}\right\} \\
 & \thinspace\thinspace\thinspace\thinspace\thinspace\thinspace\thinspace+\mathbf{e}^{\mathrm{H}}\bigtriangleup\mathbf{G}_{k}\boldsymbol{\Phi}_{k}\mathbf{\bigtriangleup\mathbf{G}}_{k}^{\mathrm{H}}\mathbf{e}\\
 & =\varepsilon_{\mathrm{h},k}^{2}\mathbf{i}_{\mathrm{h},k}^{\mathrm{H}}\boldsymbol{\Phi}_{k}\mathbf{i}_{\mathrm{h},k}+2\mathrm{Re}\left\{ \mathbf{\bigtriangleup\mathbf{h}}_{k}^{\mathrm{H}}(\boldsymbol{\Phi}_{k}\otimes\mathbf{e}^{\mathrm{T}})\mathrm{vec}(\mathbf{\bigtriangleup\mathbf{G}}_{k}^{*})\right\} \\
 & \thinspace\thinspace\thinspace\thinspace\thinspace\thinspace+\mathrm{vec}^{\mathrm{T}}(\bigtriangleup\mathbf{G}_{k})(\boldsymbol{\Phi}_{k}\otimes\mathbf{E}^{\mathrm{T}})\mathrm{vec}(\mathbf{\bigtriangleup\mathbf{G}}_{k}^{*})\\
 & =\varepsilon_{\mathrm{h},k}^{2}\mathbf{i}_{\mathrm{h},k}^{\mathrm{H}}\boldsymbol{\Phi}_{k}\mathbf{i}_{\mathrm{h},k}+2\mathrm{Re}\left\{ \varepsilon_{\mathrm{h},k}\varepsilon_{\mathrm{g},k}\mathbf{i}_{\mathrm{h},k}^{\mathrm{H}}(\boldsymbol{\Phi}_{k}\otimes\mathbf{e}^{\mathrm{T}})\mathbf{i}_{\mathrm{g},k}^{*}\right\} \\
 & \thinspace\thinspace\thinspace\thinspace\thinspace\thinspace\thinspace+\varepsilon_{\mathrm{g},k}^{2}\mathbf{i}_{\mathrm{g},k}^{\mathrm{T}}(\boldsymbol{\Phi}_{k}\otimes\mathbf{E}^{\mathrm{T}})\mathbf{i}_{\mathrm{g},k}^{*}\\
 & =\widetilde{\mathbf{i}}_{k}^{\mathrm{H}}\widetilde{\mathbf{U}}_{k}\widetilde{\mathbf{i}}_{k},
\end{align*}
where 
\[
\widetilde{\mathbf{U}}_{k}=\left[\begin{array}{cc}
\varepsilon_{\mathrm{h},k}^{2}\boldsymbol{\Phi}_{k} & \varepsilon_{\mathrm{h},k}\varepsilon_{\mathrm{g},k}(\boldsymbol{\Phi}_{k}\otimes\mathbf{e}^{\mathrm{T}})\\
\varepsilon_{\mathrm{h},k}\varepsilon_{\mathrm{g},k}(\boldsymbol{\Phi}_{k}\otimes\mathbf{e}^{\mathrm{*}}) & \varepsilon_{\mathrm{g},k}^{2}(\boldsymbol{\Phi}_{k}\otimes\mathbf{E}^{\mathrm{T}})
\end{array}\right].
\]

Denote $\widetilde{u}_{k}=(\mathbf{\widehat{h}}_{k}^{\mathrm{H}}+\mathbf{e}^{\mathrm{H}}\mathbf{\widehat{G}}_{k})\boldsymbol{\Phi}_{k}(\mathbf{\widehat{h}}_{k}+\mathbf{\widehat{G}}_{k}^{\mathrm{H}}\mathbf{e})-\sigma_{k}^{2}$,
the rate outage constraint (\ref{eq:Outage-D-1}) is then equivalent
to 
\begin{align}
 & \mathrm{Pr}\left\{ \widetilde{\mathbf{i}}_{k}^{\mathrm{H}}\widetilde{\mathbf{U}}_{k}\widetilde{\mathbf{i}}_{k}+2\mathrm{Re}\left\{ \widetilde{\mathbf{u}}_{k}^{\mathrm{H}}\widetilde{\mathbf{i}}_{k}\right\} +\widetilde{u}_{k}\geq0\right\} \geq1-\rho_{k}.\label{eq:Outage-D-2}
\end{align}

Combining Lemma \ref{Bernstein-Type Inequalities} and new auxiliary
variables $\widetilde{\mathbf{x}}=[\widetilde{x}_{1},...,\widetilde{x}_{K}]^{\mathrm{T}}$
and $\mathbf{\widetilde{y}}=[\widetilde{y}_{1},...,\widetilde{y}_{K}]^{\mathrm{T}}$,
the approximation of the data rate outage constraint of user $k$
in (\ref{eq:Outage-D-2}) is given by \begin{subequations}\label{eq:Outage-D-3}
\begin{align}
 & \mathrm{Tr}\left\{ \widetilde{\mathbf{U}}_{k}\right\} -\sqrt{2\ln(1/\rho_{k})}\widetilde{x}_{k}+\ln(\rho_{k})\widetilde{y}_{k}+\widetilde{u}_{k}\geq0,\\
 & \sqrt{||\widetilde{\mathbf{U}}_{k}||_{F}^{2}+2||\widetilde{\mathbf{u}}_{k}||^{2}}\leq\widetilde{x}_{k},\\
 & \widetilde{y}_{k}\mathbf{I}+\widetilde{\mathbf{U}}_{k}\succeq\mathbf{0},\widetilde{y}_{k}\geq0.
\end{align}
\end{subequations}

We simplify some terms in (\ref{eq:Outage-D-3}) as follows:\begin{subequations}\label{eq:Outage-D-4}
\begin{align}
 & \mathrm{Tr}\left\{ \widetilde{\mathbf{U}}_{k}\right\} =\mathrm{Tr}\Biggl\{\left[\begin{array}{c}
\varepsilon_{\mathrm{h},k}\boldsymbol{\Phi}_{k}^{a}\\
\varepsilon_{\mathrm{g},k}(\boldsymbol{\Phi}_{k}^{a}\otimes\mathbf{e}^{\mathrm{*}})
\end{array}\right]\nonumber \\
 & \hspace{1.2cm}\;\;\;\bullet\left[\begin{array}{cc}
\varepsilon_{\mathrm{h},k}\boldsymbol{\Phi}_{k}^{b} & \varepsilon_{\mathrm{g},k}(\boldsymbol{\Phi}_{k}^{b}\otimes\mathbf{e}^{\mathrm{T}})\end{array}\right]\Biggr\}\nonumber \\
 & \hspace{1.2cm}\enskip=(\varepsilon_{\mathrm{h},k}^{2}+\varepsilon_{\mathrm{g},k}^{2}M)\mathrm{Tr}\left\{ \boldsymbol{\Phi}_{k}\right\} ,\\
 & ||\widetilde{\mathbf{U}}_{k}||_{F}^{2}=(\varepsilon_{\mathrm{h},k}^{2}+\varepsilon_{\mathrm{g},k}^{2}M)^{2}||\boldsymbol{\Phi}_{k}||_{F}^{2},\\
 & ||\widetilde{\mathbf{u}}_{k}||^{2}=(\varepsilon_{\mathrm{h},k}^{2}+\varepsilon_{\mathrm{g},k}^{2}M)||(\mathbf{\widehat{h}}_{k}^{\mathrm{H}}+\mathbf{e}^{\mathrm{H}}\mathbf{\widehat{G}}_{k})\boldsymbol{\Phi}_{k}||_{2}^{2},\\
 & \widetilde{y}_{k}\mathbf{I}+\widetilde{\mathbf{U}}_{k}\succeq\mathbf{0}\Longrightarrow\widetilde{y}_{k}\mathbf{I}+(\varepsilon_{\mathrm{h},k}^{2}+\varepsilon_{\mathrm{g},k}^{2}M)\boldsymbol{\Phi}_{k}\succeq\mathbf{0}.
\end{align}
\end{subequations}The derivations of (\ref{eq:Outage-D-4}) are given
in Appendix \ref{subsec:The-proof-of-4}.

Based on the above results, Problem (\ref{Pro:min-power}) with imperfect
DCSIT and imperfect CBIUT is given by \begin{subequations}\label{Pro:min-power-D-1}
\begin{align}
\mathop{\min}\limits _{\mathbf{F},\mathbf{e},\widetilde{\mathbf{x}},\mathbf{\widetilde{y}}} & \;\;||\mathbf{F}||_{F}^{2}\\
\textrm{s.t.} & \;\;(\varepsilon_{\mathrm{h},k}^{2}+\varepsilon_{\mathrm{g},k}^{2}M)\mathrm{Tr}\left\{ \boldsymbol{\Phi}_{k}\right\} -\sqrt{2\ln(1/\rho_{k})}x_{k}\nonumber \\
 & \thinspace\thinspace\thinspace\thinspace-\ln(1/\rho_{k})y_{k}+\widetilde{u}_{k}\geq0,\forall k\in\mathcal{K}\\
 & \;\;\left\Vert \begin{array}{c}
(\varepsilon_{\mathrm{h},k}^{2}+\varepsilon_{\mathrm{g},k}^{2}M)\mathrm{vec}(\boldsymbol{\Phi}_{k})\\
\sqrt{2(\varepsilon_{\mathrm{h},k}^{2}+\varepsilon_{\mathrm{g},k}^{2}M)}\boldsymbol{\Phi}_{k}\left(\mathbf{\widehat{h}}_{k}+\mathbf{\widehat{G}}_{k}^{\mathrm{H}}\mathbf{e}\right)
\end{array}\right\Vert \leq\widetilde{x}_{k},\\
 & \thinspace\thinspace\thinspace\thinspace\thinspace\thinspace\forall k\in\mathcal{K}\\
 & \;\;\widetilde{y}_{k}\mathbf{I}+(\varepsilon_{\mathrm{h},k}^{2}+\varepsilon_{\mathrm{g},k}^{2}M)\boldsymbol{\Phi}_{k}\succeq\mathbf{0},\widetilde{y}_{k}\geq0,\forall k\in\mathcal{K}\;\;\\
 & \;\;|e_{m}|^{2}=1,\forall m\in\mathcal{M}.
\end{align}
\end{subequations}

Comparing Problem (\ref{Pro:min-power-D-1}) with Problem (\ref{Pro:min-power-1}),
we find that the former can be obtained from the latter by replacing
$\varepsilon_{\mathrm{g},k}^{2}M$ with $\varepsilon_{\mathrm{h},k}^{2}+\varepsilon_{\mathrm{g},k}^{2}M$
and $\mathbf{h}_{k}$ with $\mathbf{\widehat{h}}_{k}$. Therefore,
Problem (\ref{Pro:min-power-D-1}) can be solved by using the same
techniques as those used to solve Problem (\ref{Pro:min-power-1}).
In addition, when $M$ is large, the impact of imperfect CBIUT dominates
the performance of the system, which will be illustrated in the numerical
results later. Thus, it is significant to investigate the robust beamforming
in an IRS-aided system in which there are a large number of reflection
elements with high channel estimation error.

\section{Computational complexity}

In this section, we analyze the computational complexity of the proposed
robust transmission design methods. Since all the resulting convex
problems involving LMI, second-order cone (SOC) constraints and linear
constraints that can be solved by a standard interior point method
\cite{Ben-Tal2001convex}, we can compare the computational complexity
of different methods in terms of their worst-case runtime, the general
expression (we ignore the complexity of the linear constraints) of
which is given by 
\[
\mathcal{O}((\sum_{j=1}^{J}b_{j}+2I)^{1/2}n(n^{2}+\underbrace{n\sum_{j=1}^{J}b_{j}^{2}+\sum_{j=1}^{J}b^{3}}_{{\rm {\mathsf{\mathrm{due\thinspace\thinspace to\thinspace\thinspace LMI}}}}}+\underbrace{n\sum_{i=1}^{I}a_{i}^{2}}_{{\rm {due\thinspace\thinspace to\thinspace\thinspace SOC}}}))\text{,}
\]
where $n$ is the number of variables, $J$ is the number of LMIs
of size $b_{j}$, and $I$ is the number of SOC of size $a_{i}$.
Based on the above expression, we provide the computational complexity
per iteration of the proposed methods as follows:

1) \textit{PCU-bounded method} denotes the worst-case beamforming
design method under Scenario 1. The approximate complexity of Problem
(\ref{Pro:min-power-worst-partial-f}) is $o_{\mathbf{F}}=\mathcal{O}([K(MN+K+N+1)]^{1/2}n_{1}[n_{1}^{2}+n_{1}K((MN+1)^{2}+(K+N)^{2})+K((MN+1)^{3}+(K+N)^{3})])$
where $n_{1}=NK$, and that of Problem (\ref{Pro:min-power-5}) is
$o_{\mathbf{e}}=\mathcal{O}([K(MN+1+K)+2M]^{1/2}n_{2}[n_{2}^{2}+n_{2}K((MN+1)^{2}+K^{2})+K((MN+1)^{3}+K^{3})+n_{2}M])$
where $n_{2}=M$. Finally, the approximate complexity of PCU-bounded
method per iteration is $o_{\mathbf{F}}+o_{\mathbf{e}}$.

2) \textit{FCU-bounded method} denotes the worst-case beamforming
design method under Scenario 2. The approximate complexity of Problem
(\ref{Pro:min-power-worst-full-1}) is $o_{\mathbf{F}}+o_{\mathbf{e}}$,
where

$o_{\mathbf{F}}=\mathcal{O}([K(MN+3N+K+1)]^{1/2}n_{1}[n_{1}^{2}+n_{1}K((MN+N+1)^{2}+(K+2N)^{2})+K((MN+N+1)^{3}+(K+2N)^{2})])$
with $n_{1}=NK$, and $o_{\mathbf{e}}=\mathcal{O}([K(MN+1+K)+2M]^{1/2}n_{2}[n_{2}^{2}+n_{2}K((MN+1)^{2}+K^{2})+K((MN+1)^{3}+K^{3})+n_{2}M])$
with $n_{2}=M$.

3) \textit{PCU-statistic method} denotes the outage constrained beamforming
design method under Scenario 1. The approximate complexity of Problem
(\ref{Pro:min-power-f}) is $o_{\mathbf{F}}=\mathcal{O}([2K(N+1)]^{1/2}n_{1}[n_{1}^{2}+2n_{1}KN^{2}+2KN^{3}+nKN^{2}(N+1)^{2}])$
where $n_{1}=NK$, and that of Problem (\ref{Pro:min-power-e-1})
is $o_{\mathbf{e}}=\mathcal{O}([4K+2M]^{1/2}n_{2}[n_{2}^{2}+n_{2}(K(M^{2}+(N+1)^{2})+M)])$
where $n_{2}=M$. Finally, the approximate complexity of PCU-statistic
method per iteration is $o_{\mathbf{F}}+o_{\mathbf{e}}$.

4) \textit{FCU-statistic method} denotes the outage constrained beamforming
design method under Scenario 2. Here, the approximate complexity per
iteration is the same with the PCU-statistic method since they only
have some different coefficients.

\section{Numerical results and discussions}

\begin{figure}
\centering \includegraphics[width=3in,height=1.2in]{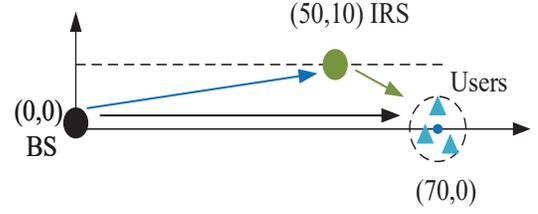}
\caption{The simulated system setup.}
\label{simulated-model} 
\end{figure}

\begin{table}
\centering \label{table} \caption{System parameters}
\begin{tabular}{|c|c|}
\hline 
Path loss exponents of BS-user link  & $\alpha_{\mathrm{BU}}=4$ \tabularnewline
\hline 
Path loss exponents of BS-IRS link  & $\alpha_{\mathrm{BI}}=2.2$ \tabularnewline
\hline 
Path loss exponents of IRS-user link  & $\alpha_{\mathrm{IU}}=2$ \tabularnewline
\hline 
Noise power  & $\sigma_{1}^{2}=...=\sigma_{K}^{2}=-80$ dBm \tabularnewline
\hline 
Convergence tolerance  & $10^{-4}$ \tabularnewline
\hline 
Maximun outage probabilities  & $\rho_{1}=...=\rho_{K}=\rho=0.05$ \tabularnewline
\hline 
\end{tabular}
\end{table}

In this section, we provide numerical results to evaluate the performance
of our proposed algorithms. The simulated system setup of our consider
network is shown in Fig. \ref{simulated-model}, in which we assume
that the BS is located at (0 m, 0 m) and the IRS is placed at (50
m, 10 m). $K$ users are randomly and uniformly distributed in a circle
centered at (70 m, 0 m) with radius of 5 m. The channel models, i.e.,
$\{\mathbf{h}_{k},\mathbf{G}_{k}\}_{\forall k\in\mathcal{K}}$, are
assumed to include large-scale fading and small-scale fading. The
large-scale fading model is expressed as $\mathrm{PL}=-\mathrm{PL}_{0}-10\alpha\log_{10}(d)$
dB, where $\alpha$ is the path loss exponent and $d$ is the link
distance in meters. $\mathrm{PL}_{0}$ denotes the pathloss at the
distance of 1 meter, which is set as $40$ dB based on the 3GPP UMi
model \cite{3Gpp-channel} with 3.5 GHz carrier frequency (i.e., carrier
frequency of 5G in China). The small-scale fading in $\{\mathbf{h}_{k},\mathbf{G}_{k}\}_{\forall k\in\mathcal{K}}$
is assumed to be Rayleigh fading distribution. For the statistical
CSI error model, the variance of $\mathrm{vec}(\bigtriangleup\mathbf{G}_{k})$
and $\bigtriangleup\mathbf{h}_{k}$ are defined as $\varepsilon_{\mathrm{g},k}^{2}=\delta_{\mathrm{g}}^{2}||\mathrm{vec}(\widehat{\mathbf{G}}_{k})||_{2}^{2}$
and $\varepsilon_{\mathrm{h},k}^{2}=\delta_{\mathrm{h}}^{2}||\widehat{\mathbf{h}}_{k}||_{2}^{2}$,
respectively. $\delta_{\mathrm{g}}\in[0,1)$ and $\delta_{\mathrm{h}}\in[0,1)$
measure the relative amount of CSI uncertainties. For the bounded
CSI error model, the radii of the uncertainty regions are set as 
\begin{align*}
\xi_{\mathrm{g},k} & =\sqrt{\frac{\varepsilon_{\mathrm{g},k}^{2}}{2}F_{2MN}^{-1}(1-\rho)},
\end{align*}
and 
\begin{align*}
\xi_{\mathrm{h},k} & =\sqrt{\frac{\varepsilon_{\mathrm{h},k}^{2}}{2}F_{2N}^{-1}(1-\rho)},
\end{align*}
where $F_{2MN}^{-1}(\cdot)$ and $F_{2N}^{-1}(\cdot)$ denote the
inverse cumulative distribution function (CDF) of the Chi-square distribution
with degrees of freedom equal to $2MN$ and $2N$, respectively. According
to \cite{W-K-MA2014}, the above bounded CSI error model provides
a fair comparison between the performance of the worst-case robust
design and the outage constrained robust design. In addition, the
target rates of all users are assumed to be the same, i.e., $R_{1}=...=R_{K}=R$
and the fixed simulation settings for our simulations are given in
Table I.

\begin{figure}
\centering \includegraphics[width=3.4in,height=2.6in]{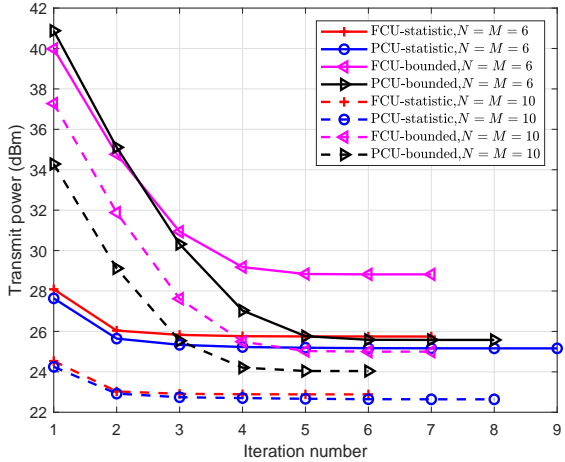}
\caption{Transmit power versus the iteration numbers of different algorithms,
when $K=3$ and $\{\delta_{\mathrm{g}},\delta_{\mathrm{h}}\}=\{0.01,0.02\}$.}
\label{Convergence} 
\end{figure}

Fig. \ref{Convergence} illustrates the convergence behavior of the
proposed four algorithms. Here, the minimum rate is set as $R=2$
bit/s/Hz, and the channel uncertainty levels are chosen as $\{\delta_{\mathrm{g}},\delta_{\mathrm{h}}\}=\{0.01,0.02\}$.
It is observed that all algorithms converge rapidly and 10 iterations
are sufficient for the algorithms to converge. It also shows that
the convergence speed increases with the number of antennas. In addition,
the algorithms under the statistical error model converge faster than
those under the bounded error model.

\begin{figure}
\centering \includegraphics[width=3.4in,height=2.6in]{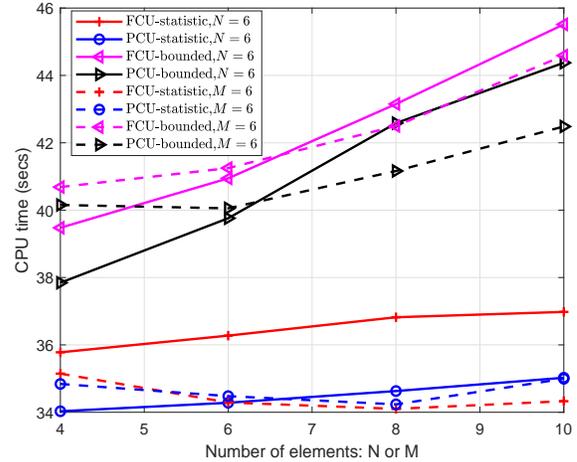} \caption{Average CPU time versus the number of antenna elements at the IRS
$M$ and at the BS $N$, when $K=2$ and $\{\delta_{\mathrm{g}},\delta_{\mathrm{h}}\}=\{0.01,0.02\}$.}
\label{time} 
\end{figure}

Fig. \ref{time} compares the average (central processing unit) CPU
running time of the proposed algorithms versus the numbers of antenna
elements at the BS and/or reflection elements at the IRS. The results
are obtained by using a computer with a 1.99 GHz i7-8550U CPU and
16 GB RAM. Here, we set $K=2$, $R=2$ bit/s/Hz, and $\{\delta_{\mathrm{g}},\delta_{\mathrm{h}}\}=\{0.01,0.02\}$.
Firstly, it is observed that the robust algorithms under the statistical
CSI error model require much less CPU running time than those under
the bounded CSI error model. This is due to the fact that there are
some large-dimensional LMIs that increase the computational complexity
of the worst-case algorithms. Secondly, the FCU-bounded algorithm
requires more CPU time than the PCU-bounded algorithm because the
DCSIT error $\bigtriangleup\mathbf{h}_{k}$ increases the dimension
of the LMIs. Thirdly, when $M=6$, the CPU running time of the outage
constrained algorithm under two scenarios is similar due to the fact
that no additional complexity is introduced by considering the additional
DCSIT error.

\begin{figure}
\centering \includegraphics[width=3.4in,height=2.6in]{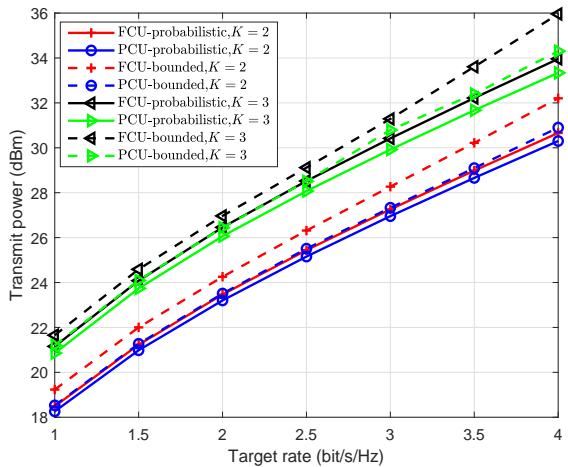}
\caption{Transmit power versus the target rate $R$ under $N=M=6$ and $\{\delta_{\mathrm{g}},\delta_{\mathrm{h}}\}=\{0.01,0.02\}$.}
\label{power-vs-rate} 
\end{figure}

Fig. \ref{power-vs-rate} shows the minimum transmit power of the
IRS-aided communication system versus the target rate requirements
of users under various CSI error models. Some system parameters are
set as $N=M=6$, $K=\{2,3\}$, $\{\delta_{\mathrm{g}},\delta_{\mathrm{h}}\}=\{0.01,0.02\}$.
It is seen that the minimum transmit power increases with the target
rate for both channel uncertainty scenarios and both CSI error models.
In addition, it is also observed that the minimum transmit power of
the worst-case robust design algorithms is larger than that of the
outage constrained robust design algorithms. This is due to the fact
that the worst-case optimization is the most conservative robust design,
which requires more transmit power with the aim of ensuring that the
achievable rate of each user meets the target rate requirement for
the worst-case CSI error realization.

\begin{figure}
\centering \subfigure[Feasibility rate]{\includegraphics[width=3.4in,height=2.6in]{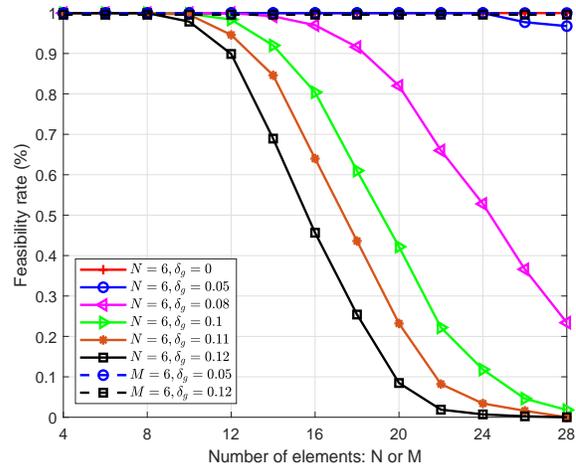}}

\centering \subfigure[Transmit power]{\includegraphics[width=3.4in,height=2.6in]{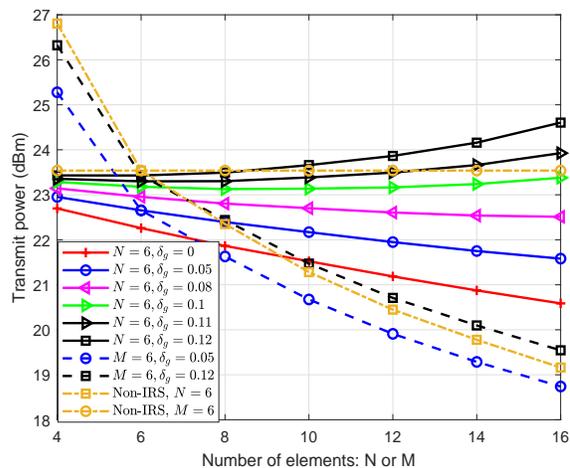}}

\caption{Feasibility rate and transmit power versus the number of antenna elements
under the PCU scenario, when $K=2$.}
\label{PCU-MN} 
\end{figure}

In the following, we study the impact of the accuracy of the CSI on
the system performance. We adopt outage constrained robust beamforming
design algorithms since the computational complexity of the worst-case
robust beamforming design algorithms is unacceptable at large numbers
of antennas.

Fig. \ref{PCU-MN} shows the feasibility rate and the minimum transmit
power versus $N$ or $M$ when only the CBIUT is imperfect, i.e.,
$\delta_{\mathrm{h}}=0$. We assume there are $K=2$ users with $R=2$
bit/s/Hz. The feasibility rate is defined as the ratio of the number
of feasible channel realizations to the total number of channel realizations,
where the feasible channel realization means that there exists a feasible
solution to the outage constrained problem in (\ref{Pro:min-power})
with this channel realization. An interesting phenomenon can be observed
from Fig. 6 (a). When fixing the number of transmit antennas $N$,
the feasibility rate decreases rapidly with the number of phase shifters
at a high level of channel uncertainty ($\delta_{\mathrm{g}}\geq0.08$).
By contrast, when fixing the number of phase shifters $M$ the feasibility
rate keeps stable for different numbers of antennas even at a high
level of channel uncertainty.

Based on the observations of Fig. \ref{PCU-MN}(a), we further examine
the minimum transmit power consumption of different channel uncertainty
levels in Fig. \ref{PCU-MN}(b) with a benchmark scheme without IRS.
Fig. \ref{PCU-MN}(b) is generated based on the channel realizations
for which the feasible solutions can be obtained at $N=16$ or $M=16$.

We first study the case with fixed number of transmit antennas $N=6$.
In Fig. \ref{PCU-MN}(b), the case with $\delta_{\mathrm{g}}=0$ can
be regarded as the perfect CBIUT case, and its minimum transmit power
decreases with the number of the reflection elements. This trend is
consistent with that of Fig. 4 in \cite{qingqing2019}. The minimum
transmit power consumption values under small values of $\delta_{\mathrm{g}}$,
e.g., $\delta_{\mathrm{g}}=\{0.05,0.08\}$, also decrease with the
number of the reflection elements, and are higher than that of the
perfect CBIUT case. The reason is that the BS needs to consume more
power to compensate for the rate loss caused by the CBIUT error. However,
when $\delta_{\mathrm{g}}$ increases to $0.1$ or larger, transmit
power consumption starts to increase with the number of reflection
elements. The reason is that increasing the number of reflection elements
cannot only reduce the transmit power due to its increased beamforming
gain, but also increase the channel estimation error that more transmit
power is required to compensate for the channel errors. Therefore,
when the CBIUT error is small, the benefits brought by the increase
of $M$, outweighs its drawbacks, and vice versa. As a result, the
number of IRS reflection elements should be carefully chosen, and
the accuracy of the CBIUT estimation is crucial to reap the benefits
offered by the IRS.

On the other hand, for the case with a fixed number of reflection
elements, the transmit power consumption values decrease with the
number of antennas at the BS even when the CBIUT error is high as
$\delta_{\mathrm{g}}=0.12$. The reason is that when the number of
antennas is large, more degrees of freedom can be exploited to optimize
the active beamforming vector at the BS to compensate for the channel
estimation error. Finally, compared with the system without IRS, the
IRS may lose its performance gain advantage under high CBIUT error.

\begin{figure}
\centering \subfigure[Feasibility rate]{\includegraphics[width=3.4in,height=2.6in]{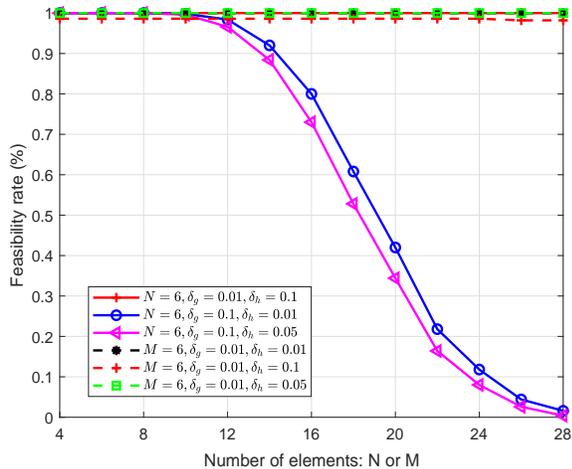}}

\centering \subfigure[Transmit power]{\includegraphics[width=3.4in,height=2.6in]{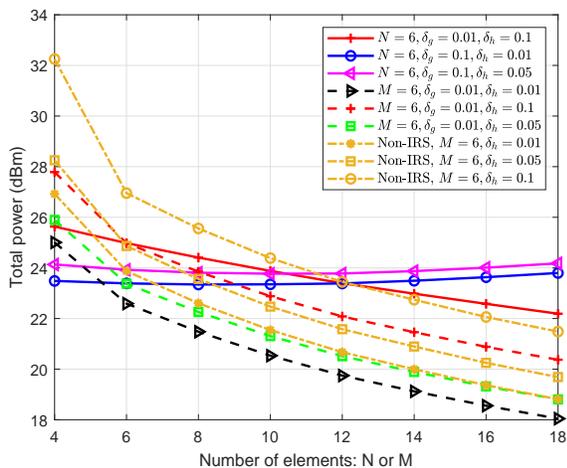}}

\caption{Feasibility rate and transmit power versus the number of antenna elements
under the FCU scenario, when $K=2$.}
\label{FCU-MN} 
\end{figure}

Fig. \ref{FCU-MN} shows the feasibility rate and the minimum transmit
power versus $M$ or $N$ when both the DCSIT and the CBIUT are imperfect.
The simulation parameters are the same as those in Fig. \ref{PCU-MN}.
Fig. \ref{FCU-MN}(a) shows that when $\delta_{\mathrm{g}}$ is low,
the feasibility rates achieved by various cases are always high. In
addition, from Fig. \ref{FCU-MN}(b), we find that the increase of
the number of antennas at the BS is effective in reducing the transmit
power consumption, which is not affected by the DCSIT error $\delta_{\mathrm{h}}$
(see curves $M=6,\delta_{\mathrm{g}}=0.01,\delta_{\mathrm{h}}=\{0.01,0.05,0.1\}$).

\section{Conclusions}

In this work, we investigated robust beamforming designs under imperfect
CBIUT for the IRS-aided MU-MISO system. Our aim was to minimize the
transmit power subject to the worst-case rate constraints under the
bounded CSI error model and the rate outage probability constraints
under the statistical CSI error model. The CSI uncertainties under
the bounded CSI error model were addressed by applying the S-procedure,
and those under the statistical CSI error model were tackled by using
the Bernstein-Type Inequality. The reformulated problems were efficiently
solved under the AO framework. It is shown that the performance in
terms of the minimum achievable transmit power, convergence and complexity
under the statistical CSI error model is higher than that under the
bounded CSI error model. The number of elements on the IRS may have
a negative impact on system performance when the CBIUT error is large.
This conclusion provides an engineering insight for the careful selection
of the size of the IRS. In the end, this work provides a framework
of robust transmission design in a simple single-cell multiuser scenario.
The more complicated scenarios, such as the IRSs-assisted full-duplex
communication systems, IRS-aided energy efficiency systems and IRS-aided
physical layer security systems, will be studied as our future work.
Furthermore, the robustness of the IRS in millimeter wave system under
a geometric channel model is also worth studying.

\appendices{}

\section{The proof of Lemma \ref{lower-bound}\label{subsec:The-proof-of-1}}

Let $x$ be a complex scalar variable, we have the first-order Taylor
inequality 
\begin{equation}
\left|x\right|^{2}\geq2\mathrm{Re}\left\{ x^{*,(n)}x\right\} -x^{*,(n)}x^{(n)},\label{eq:df}
\end{equation}
for any fixed point $x^{(n)}$. By replacing $x$ and $x^{(n)}$ in
(\ref{eq:df}) with $(\mathbf{h}_{k}^{\mathrm{H}}+\mathbf{e}^{\mathrm{H}}\mathbf{G}_{k}){\bf f}_{k}$
and $(\mathbf{h}_{k}^{\mathrm{H}}+\mathbf{e}^{(n),\mathrm{H}}\mathbf{G}_{k}){\bf f}_{k}^{(n)}$,
respectively, we have 
\begin{align}
 & \left|\left(\mathbf{h}_{k}^{\mathrm{H}}+\mathbf{e}^{\mathrm{H}}\mathbf{G}_{k}\right){\bf f}_{k}\right|^{2}\nonumber \\
 & \geq2\mathrm{Re}\left\{ \left(\mathbf{h}_{k}^{\mathrm{H}}+\mathbf{e}^{(n),\mathrm{H}}\mathbf{G}_{k}\right){\bf f}_{k}^{(n)}\mathbf{f}_{k}^{\mathrm{H}}\left(\mathbf{h}_{k}+\mathbf{G}_{k}^{\mathrm{H}}\mathbf{e}\right)\right\} \nonumber \\
 & \ \ \ -\left(\mathbf{h}_{k}^{\mathrm{H}}+\mathbf{e}^{(n),\mathrm{H}}\mathbf{G}_{k}\right){\bf f}_{k}^{(n)}\mathbf{f}_{k}^{(n),\mathrm{H}}\left(\mathbf{h}_{k}+\mathbf{G}_{k}^{\mathrm{H}}\mathbf{e}^{(n)}\right).\label{eq:gdg}
\end{align}

By plugging $\mathbf{G}_{k}=\widehat{\mathbf{G}}_{k}+\bigtriangleup\mathbf{G}_{k}$
into the right hand side of (\ref{eq:gdg}) and expanding it by using
mathematical transformations, i.e., $\mathrm{Tr}(\mathbf{A}^{\mathrm{H}}\mathbf{B})=\mathrm{vec}^{\mathrm{H}}(\mathbf{A})\mathrm{vec}(\mathbf{B})$
and $\mathrm{Tr}(\mathbf{A}\mathbf{B}\mathbf{C}\mathbf{D})=(\mathrm{vec}^{\mathrm{T}}(\mathbf{D}))^{\mathrm{T}}(\mathbf{C}^{\mathrm{T}}\otimes\mathbf{A})\mathrm{vec}(\mathbf{B})$
\cite{Xinda2017}, we can obtained (\ref{eq:lower-bound-1}).

Hence, the proof is completed.

\section{The proof of Lemma \ref{lower-bound-1}\label{subsec:The-proof-of-2}}

The lower bound of (\ref{eq:lower-bound-full}) can also be derived
from (\ref{eq:gdg}) under the full channel uncertainty. In particular,
we insert $\mathbf{h}_{k}=\widehat{\mathbf{h}}_{k}+\bigtriangleup\mathbf{h}_{k}$
and $\mathbf{G}_{k}=\widehat{\mathbf{G}}_{k}+\bigtriangleup\mathbf{G}_{k}$
into the first term on the right hand side of (\ref{eq:gdg}), and
then get (\ref{eq:gg}) at the top of the next page. 
\begin{figure*}
\begin{align}
 & \left[(\widehat{\mathbf{h}}_{k}^{\mathrm{H}}+\bigtriangleup\mathbf{h}_{k}^{\mathrm{H}})+\mathbf{e}^{(n),\mathrm{H}}(\widehat{\mathbf{G}}_{k}+\bigtriangleup\mathbf{G}_{k})\right]{\bf f}_{k}^{(n)}\mathbf{f}_{k}^{\mathrm{H}}\left[(\widehat{\mathbf{h}}_{k}+\bigtriangleup\mathbf{h}_{k})+(\widehat{\mathbf{G}}_{k}^{\mathrm{H}}+\bigtriangleup\mathbf{G}_{k}^{\mathrm{H}})\mathbf{e}\right]\nonumber \\
= & (\mathbf{\widehat{h}}_{k}^{\mathrm{H}}+\mathbf{e}^{(n),\mathrm{H}}\mathbf{\widehat{G}}_{k}){\bf f}_{k}^{(n)}\mathbf{f}_{k}^{\mathrm{H}}(\mathbf{\widehat{h}}_{k}+\mathbf{\widehat{G}}_{k}^{\mathrm{H}}\mathbf{e})+(\mathbf{\widehat{h}}_{k}^{\mathrm{H}}+\mathbf{e}^{(n),\mathrm{H}}\mathbf{\widehat{G}}_{k}){\bf f}_{k}^{(n)}\mathbf{f}_{k}^{\mathrm{H}}(\bigtriangleup\mathbf{h}_{k}+\mathbf{\bigtriangleup\mathbf{G}}_{k}^{\mathrm{H}}\mathbf{e})\nonumber \\
 & +(\mathbf{\bigtriangleup\mathbf{h}}_{k}^{\mathrm{H}}+\mathbf{e}^{(n),\mathrm{H}}\bigtriangleup\mathbf{G}_{k}){\bf f}_{k}^{(n)}\mathbf{f}_{k}^{\mathrm{H}}(\mathbf{\widehat{h}}_{k}+\mathbf{\widehat{G}}_{k}^{\mathrm{H}}\mathbf{e})+(\mathbf{\bigtriangleup\mathbf{h}}_{k}^{\mathrm{H}}+\mathbf{e}^{(n),\mathrm{H}}\bigtriangleup\mathbf{G}_{k}){\bf f}_{k}^{(n)}\mathbf{f}_{k}^{\mathrm{H}}(\bigtriangleup\mathbf{h}_{k}+\mathbf{\bigtriangleup\mathbf{G}}_{k}^{\mathrm{H}}\mathbf{e})\nonumber \\
= & (\mathbf{\widehat{h}}_{k}^{\mathrm{H}}+\mathbf{e}^{(n),\mathrm{H}}\mathbf{\widehat{G}}_{k}){\bf f}_{k}^{(n)}\mathbf{f}_{k}^{\mathrm{H}}(\mathbf{\widehat{h}}_{k}+\mathbf{\widehat{G}}_{k}^{\mathrm{H}}\mathbf{e})+(\mathbf{\widehat{h}}_{k}^{\mathrm{H}}+\mathbf{e}^{(n),\mathrm{H}}\mathbf{\widehat{G}}_{k}){\bf f}_{k}^{(n)}\mathbf{f}_{k}^{\mathrm{H}}\bigtriangleup\mathbf{h}_{k}\nonumber \\
 & +\mathrm{vec}^{\mathrm{H}}(\mathbf{\bigtriangleup\mathbf{G}}_{k})\mathrm{vec}(\mathbf{e}(\mathbf{\widehat{h}}_{k}^{\mathrm{H}}+\mathbf{e}^{(n),\mathrm{H}}\mathbf{\widehat{G}}_{k}){\bf f}_{k}^{(n)}\mathbf{f}_{k}^{\mathrm{H}})+\mathbf{\bigtriangleup\mathbf{h}}_{k}^{\mathrm{H}}{\bf f}_{k}^{(n)}\mathbf{f}_{k}^{\mathrm{H}}(\mathbf{\widehat{h}}_{k}+\mathbf{\widehat{G}}_{k}^{\mathrm{H}}\mathbf{e})+\mathbf{\bigtriangleup\mathbf{h}}_{k}^{\mathrm{H}}{\bf f}_{k}^{(n)}\mathbf{f}_{k}^{\mathrm{H}}\bigtriangleup\mathbf{h}_{k}\nonumber \\
 & +\mathrm{vec}^{\mathrm{H}}(\mathbf{e}^{(n)}(\mathbf{\widehat{h}}_{k}+\mathbf{e}^{\mathrm{H}}\mathbf{\widehat{G}}_{k}^{\mathrm{H}}){\bf f}_{k}\mathbf{f}_{k}^{(n),\mathrm{H}})\mathrm{vec}(\mathbf{\bigtriangleup\mathbf{G}}_{k})\ +\mathrm{vec}^{\mathrm{H}}(\mathbf{\bigtriangleup\mathbf{G}}_{k})(\mathbf{f}_{k}^{*}{\bf f}_{k}^{(n),\mathrm{T}}\otimes\mathbf{e})\mathbf{\bigtriangleup\mathbf{h}}_{k}^{*}\nonumber \\
 & +\bigtriangleup\mathbf{h}_{k}^{\mathrm{T}}(\mathbf{f}_{k}^{*}{\bf f}_{k}^{(n),\mathrm{T}}\otimes\mathbf{e}^{(n),\mathrm{H}})\mathrm{vec}(\mathbf{\bigtriangleup\mathbf{G}}_{k})+\mathrm{vec}^{\mathrm{H}}(\mathbf{\bigtriangleup\mathbf{G}}_{k})(\mathbf{f}_{k}^{*}{\bf f}_{k}^{(n),\mathrm{T}}\otimes\mathbf{e}\mathbf{e}^{(n),\mathrm{H}})\mathrm{vec}(\mathbf{\bigtriangleup\mathbf{G}}_{k})\nonumber \\
= & \widetilde{\mathbf{i}}_{k}^{\mathrm{H}}\mathbf{D}_{k}\widetilde{\mathbf{i}}_{k}+\mathbf{d}_{1,k}^{\mathrm{H}}\widetilde{\mathbf{i}}_{k}+\widetilde{\mathbf{i}}_{k}^{\mathrm{H}}\mathbf{d}_{2,k}+d_{k}.\label{eq:gg}
\end{align}
\hrule 
\end{figure*}

With the similar mathematical transformations, the remaining two terms
on the right hand side of (\ref{eq:gdg}) under the full channel uncertainty
can be expressed as 
\begin{align}
 & (\mathbf{h}_{k}^{\mathrm{H}}+\mathbf{e}^{\mathrm{H}}\mathbf{G}_{k}){\bf f}_{k}\mathbf{f}_{k}^{(n),\mathrm{H}}(\mathbf{h}_{k}+\mathbf{G}_{k}^{\mathrm{H}}\mathbf{e}^{(n)})\nonumber \\
= & \widetilde{\mathbf{i}}_{k}^{\mathrm{H}}\mathbf{D}_{k}^{\mathrm{H}}\widetilde{\mathbf{i}}_{k}+\mathbf{d}_{2,k}^{\mathrm{H}}\widetilde{\mathbf{i}}_{k}+\widetilde{\mathbf{i}}_{k}^{\mathrm{H}}\mathbf{d}_{1,k}+d_{k}^{*}\nonumber \\
 & +(\mathbf{h}_{k}^{\mathrm{H}}+\mathbf{e}^{(n),\mathrm{H}}\mathbf{G}_{k}){\bf f}_{k}^{(n)}\mathbf{f}_{k}^{(n),\mathrm{H}}(\mathbf{h}_{k}+\mathbf{G}_{k}^{\mathrm{H}}\mathbf{e}^{(n)})\nonumber \\
= & \widetilde{\mathbf{i}}_{k}^{\mathrm{H}}\mathbf{Z}_{k}\widetilde{\mathbf{i}}_{k}+\mathbf{z}_{k}^{\mathrm{H}}\widetilde{\mathbf{i}}_{k}+\widetilde{\mathbf{i}}_{k}^{\mathrm{H}}\mathbf{z}_{k}+z_{k}.
\end{align}

Hence, the proof is completed.

\section{The proof of Theorem \ref{Theorem-1}\label{subsec:The-proof-of-rank-1}}

Denote by $\widehat{\boldsymbol{\Gamma}}^{\star}=[\boldsymbol{\widehat{\Gamma}}_{1}^{\star},...,\boldsymbol{\widehat{\Gamma}}_{K}^{\star}]$
the optimal solution of the relaxed version of Problem (\ref{Pro:min-power-f})
and define the projection matrices as $\mathbf{P}_{k}={\widehat{\boldsymbol{\Gamma}}_{k}^{\star\frac{1}{2}}\mathbf{\widehat{h}}_{k}\mathbf{\widehat{h}}_{k}^{\mathrm{H}}\widehat{\boldsymbol{\Gamma}}_{k}^{\star\frac{1}{2}}}/||\widehat{\boldsymbol{\Gamma}}_{k}^{\star\frac{1}{2}}\mathbf{\widehat{h}}_{k}||^{2},\forall k\in\mathcal{K}$,
where $\mathbf{\widehat{h}}_{k}=\left(\mathbf{h}_{k}+\mathbf{G}_{k}^{\mathrm{H}}\mathbf{e}\right)$.
Then, we construct a rank-one solution $\boldsymbol{\widetilde{\Gamma}}^{\star}=[\boldsymbol{\widetilde{\Gamma}}_{1}^{\star},...,\boldsymbol{\widetilde{\Gamma}}_{K}^{\star}]$,
each sub-matrix of which is given by 
\begin{equation}
\boldsymbol{\widetilde{\Gamma}}_{k}^{\star}=\widehat{\boldsymbol{\Gamma}}_{k}^{\star\frac{1}{2}}\mathbf{P}_{k}\widehat{\boldsymbol{\Gamma}}_{k}^{\star\frac{1}{2}}.\label{eq:new-solution}
\end{equation}

Firstly, we check the objective value of Problem (\ref{Pro:min-power-f})
with solution $\boldsymbol{\widetilde{\Gamma}}^{\star}$: 
\begin{align}
 & \sum_{k=1}^{K}\mathrm{Tr}\left\{ \boldsymbol{\widetilde{\Gamma}}_{k}^{\star}\right\} -\sum_{k=1}^{K}\mathrm{Tr}\left\{ \widehat{\boldsymbol{\Gamma}}_{k}^{\star}\right\}  & =\sum_{k=1}^{K}\mathrm{Tr}\left\{ \widehat{\boldsymbol{\Gamma}}_{k}^{\star\frac{1}{2}}(\mathbf{P}_{k}-\mathbf{I})\widehat{\boldsymbol{\Gamma}}_{k}^{\star\frac{1}{2}}\right\} \nonumber \\
 & \leq0,\label{eq:obj}
\end{align}
which means that the objective value acheived by using the solution
$\boldsymbol{\widetilde{\Gamma}}^{\star}$ is no more than that generated
from the optimal solution $\widehat{\boldsymbol{\Gamma}}^{\star}$.

Then, since it is computationally intractable to check whether the
constructed solution satisfies the constraints (\ref{eq:cons-1-1})-(\ref{eq:sdf-1})
directly, we instead consider the constraint (\ref{eq:min-power-cons1})
in the original Problem (\ref{Pro:min-power}). Specifically, from
(\ref{eq:Outage-1}), we have 
\begin{equation}
\frac{\mathbf{\widehat{h}}_{k}^{\mathrm{H}}\boldsymbol{\widetilde{\Gamma}}_{k}^{\star}\mathbf{\widehat{h}}_{k}}{(2^{R_{k}}-1)}=\frac{|\mathbf{\widehat{h}}_{k}^{\mathrm{H}}\widehat{\boldsymbol{\Gamma}}_{k}^{\star}\mathbf{\widehat{h}}_{k}|^{2}}{||\widehat{\boldsymbol{\Gamma}}_{k}^{\star\frac{1}{2}}\mathbf{\widehat{h}}_{k}||^{2}(2^{R_{k}}-1)}=\frac{\mathbf{\widehat{h}}_{k}^{\mathrm{H}}\widehat{\boldsymbol{\Gamma}}_{k}^{\star}\mathbf{\widehat{h}}_{k}}{(2^{R_{k}}-1)},\label{eq:d1}
\end{equation}
as well as 
\begin{align}
\mathbf{\widehat{h}}_{k}^{\mathrm{H}}\boldsymbol{\widetilde{\Gamma}}_{i}^{\star}\mathbf{\widehat{h}}_{k} & =\mathbf{\widehat{h}}_{i}^{\mathrm{H}}\widehat{\boldsymbol{\Gamma}}_{i}^{\star\frac{1}{2}}\frac{\widehat{\boldsymbol{\Gamma}}_{i}^{\star\frac{1}{2}}\mathbf{\widehat{h}}_{k}\mathbf{\widehat{h}}_{k}^{\mathrm{H}}\widehat{\boldsymbol{\Gamma}}_{i}^{\star\frac{1}{2}}}{||\widehat{\boldsymbol{\Gamma}}_{i}^{\star\frac{1}{2}}\mathbf{\widehat{h}}_{i}||^{2}}\widehat{\boldsymbol{\Gamma}}_{i}^{\star\frac{1}{2}}\mathbf{\widehat{h}}_{i}\nonumber \\
 & \leq\lambda_{max}\left(\widehat{\boldsymbol{\Gamma}}_{i}^{\star\frac{1}{2}}\mathbf{\widehat{h}}_{k}\mathbf{\widehat{h}}_{k}^{\mathrm{H}}\widehat{\boldsymbol{\Gamma}}_{i}^{\star\frac{1}{2}}\right)=\mathbf{\widehat{h}}_{k}^{\mathrm{H}}\widehat{\boldsymbol{\Gamma}}_{i}^{\star}\mathbf{\widehat{h}}_{k}.\label{eq:d2}
\end{align}
Combining (\ref{eq:d1}) with (\ref{eq:d2}), we have 
\begin{align}
 & \mathbf{\widehat{h}}_{k}^{\mathrm{H}}[\boldsymbol{\widetilde{\Gamma}}_{k}^{\star}/(2^{R_{k}}-1)-\sum_{i\neq k}^{K}\boldsymbol{\widetilde{\Gamma}}_{k}^{\star}]\mathbf{\widehat{h}}_{k}\nonumber \\
 & \geq\mathbf{\widehat{h}}_{k}^{\mathrm{H}}[\widehat{\boldsymbol{\Gamma}}_{k}^{\star}/(2^{R_{k}}-1)-\sum_{i\neq k}^{K}\widehat{\boldsymbol{\Gamma}}_{i}^{\star}]\mathbf{\widehat{h}}_{k},\label{eq:vvv}
\end{align}
which implies that the constructed solution $\boldsymbol{\widetilde{\Gamma}}^{\star}$
satisfies constraint (\ref{eq:min-power-cons1}) and then satisfies
constraints (\ref{eq:cons-1-1})-(\ref{eq:sdf-1}).

With (\ref{eq:obj}) and (\ref{eq:vvv}), we conclude that $\boldsymbol{\widetilde{\Gamma}}^{\star}$
is also a feasible solution of the relaxed version of Problem (\ref{Pro:min-power-f})
with rank one.

Hence, the proof is completed.

\section{Derivation of (\ref{eq:Outage-D-4}) \label{subsec:The-proof-of-4}}

Denote 
\begin{align*}
\boldsymbol{\Phi}_{k} & ={\bf f}_{k}\mathbf{f}_{k}^{\mathrm{H}}/(2^{R_{k}}-1)-{\bf F}_{-k}\mathbf{F}_{-k}^{\mathrm{H}}\\
 & =\boldsymbol{\Phi}_{k}^{a}\boldsymbol{\Phi}_{k}^{b},
\end{align*}
where 
\begin{align*}
\boldsymbol{\Phi}_{k}^{a} & =\left[\sqrt{1/(1-2^{R_{k}})}\begin{array}{cc}
{\bf f}_{k} & {\color{red}{\color{blue}j}}{\bf F}_{-k}\end{array}\right],\\
\boldsymbol{\Phi}_{k}^{b} & =\left[\begin{array}{c}
\sqrt{1/(1-2^{R_{k}})}\mathbf{f}_{k}^{\mathrm{H}}\\
{\color{red}{\color{blue}j}}\mathbf{F}_{-k}^{\mathrm{H}}
\end{array}\right].
\end{align*}

Then
\begin{align*}
 & \mathrm{Tr}\left\{ \widetilde{\mathbf{U}}_{k}\right\} \\
 & =\mathrm{Tr}\Biggl\{\left[\begin{array}{c}
\varepsilon_{\mathrm{h},k}(\boldsymbol{\Phi}_{k}^{a}\otimes1)\\
\varepsilon_{\mathrm{g},k}(\boldsymbol{\Phi}_{k}^{a}\otimes\mathbf{e}^{\mathrm{*}})
\end{array}\right]\\
 & \bullet\left[\begin{array}{cc}
\varepsilon_{\mathrm{h},k}(\boldsymbol{\Phi}_{k}^{b}\otimes1) & \varepsilon_{\mathrm{g},k}(\boldsymbol{\Phi}_{k}^{b}\otimes\mathbf{e}^{\mathrm{T}})\end{array}\right]\Biggr\}\\
 & =\mathrm{Tr}\Biggl\{\left[\begin{array}{cc}
\varepsilon_{\mathrm{h},k}(\boldsymbol{\Phi}_{k}^{b}\otimes1) & \varepsilon_{\mathrm{g},k}(\boldsymbol{\Phi}_{k}^{b}\otimes\mathbf{e}^{\mathrm{T}})\end{array}\right]\\
 & \bullet\left[\begin{array}{c}
\varepsilon_{\mathrm{h},k}(\boldsymbol{\Phi}_{k}^{a}\otimes1)\\
\varepsilon_{\mathrm{g},k}(\boldsymbol{\Phi}_{k}^{a}\otimes\mathbf{e}^{\mathrm{*}})
\end{array}\right]\Biggr\}\\
 & =\mathrm{Tr}\Biggl\{\varepsilon_{\mathrm{h},k}^{2}(\boldsymbol{\Phi}_{k}^{b}\otimes1)(\boldsymbol{\Phi}_{k}^{a}\otimes1)+\varepsilon_{\mathrm{g},k}^{2}(\boldsymbol{\Phi}_{k}^{b}\otimes\mathbf{e}^{\mathrm{T}})(\boldsymbol{\Phi}_{k}^{a}\otimes\mathbf{e}^{\mathrm{*}})\Biggr\}\\
 & =\mathrm{Tr}\left\{ \left[\begin{array}{cc}
\varepsilon_{\mathrm{h},k}^{2}\boldsymbol{\Phi}_{k}^{b}\boldsymbol{\Phi}_{k}^{a}+\varepsilon_{\mathrm{g},k}^{2} & M\boldsymbol{\Phi}_{k}^{b}\end{array}\boldsymbol{\Phi}_{k}^{a}\right]\right\} \\
 & =\mathrm{Tr}\left\{ \left[\begin{array}{cc}
\varepsilon_{\mathrm{h},k}^{2}\boldsymbol{\Phi}_{k}^{a}\boldsymbol{\Phi}_{k}^{b}+\varepsilon_{\mathrm{g},k}^{2} & M\boldsymbol{\Phi}_{k}^{a}\boldsymbol{\Phi}_{k}^{b}\end{array}\right]\right\} \\
 & =\mathrm{Tr}\left\{ \left[\begin{array}{cc}
\varepsilon_{\mathrm{h},k}^{2}\boldsymbol{\Phi}_{k}+\varepsilon_{\mathrm{g},k}^{2} & M\boldsymbol{\Phi}_{k}\end{array}\right]\right\} \\
 & =(\varepsilon_{\mathrm{h},k}^{2}+\varepsilon_{\mathrm{g},k}^{2}M)\mathrm{Tr}\left\{ \boldsymbol{\Phi}_{k}\right\} 
\end{align*}
using property $\mathrm{Tr}\left\{ \mathbf{A}\mathbf{B}\right\} =\mathrm{Tr}\left\{ \mathbf{B}\mathbf{A}\right\} $.

Denoting $\lambda_{\textrm{nonzero}}\left(\mathbf{X}\right)$ the
non-zero eignvalue of $\mathbf{X}$ and using property $\lambda_{\textrm{nonzero}}\left(\mathbf{A}\mathbf{B}\right)=\lambda_{\textrm{nonzero}}\left(\mathbf{B}\mathbf{A}\right)$,
we have
\begin{align*}
 & \lambda_{\textrm{nonzero}}\left(\widetilde{\mathbf{U}}_{k}\right)\\
 & =\lambda_{\textrm{nonzero}}\Bigl(\left[\begin{array}{c}
\varepsilon_{\mathrm{h},k}(\boldsymbol{\Phi}_{k}^{a}\otimes1)\\
\varepsilon_{\mathrm{g},k}(\boldsymbol{\Phi}_{k}^{a}\otimes\mathbf{e}^{\mathrm{*}})
\end{array}\right]\\
 & \bullet\left[\begin{array}{cc}
\varepsilon_{\mathrm{h},k}(\boldsymbol{\Phi}_{k}^{b}\otimes1) & \varepsilon_{\mathrm{g},k}(\boldsymbol{\Phi}_{k}^{b}\otimes\mathbf{e}^{\mathrm{T}})\end{array}\right]\Bigl)\\
 & =\lambda_{\textrm{nonzero}}\Bigl(\left[\begin{array}{cc}
\varepsilon_{\mathrm{h},k}(\boldsymbol{\Phi}_{k}^{b}\otimes1) & \varepsilon_{\mathrm{g},k}(\boldsymbol{\Phi}_{k}^{b}\otimes\mathbf{e}^{\mathrm{T}})\end{array}\right]\\
 & \bullet\left[\begin{array}{c}
\varepsilon_{\mathrm{h},k}(\boldsymbol{\Phi}_{k}^{a}\otimes1)\\
\varepsilon_{\mathrm{g},k}(\boldsymbol{\Phi}_{k}^{a}\otimes\mathbf{e}^{\mathrm{*}})
\end{array}\right]\Bigl)\\
 & =\lambda_{\textrm{nonzero}}\left(\varepsilon_{\mathrm{h},k}^{2}(\boldsymbol{\Phi}_{k}^{b}\otimes1)(\boldsymbol{\Phi}_{k}^{a}\otimes1)+\varepsilon_{\mathrm{g},k}^{2}(\boldsymbol{\Phi}_{k}^{b}\otimes\mathbf{e}^{\mathrm{T}})(\boldsymbol{\Phi}_{k}^{a}\otimes\mathbf{e}^{\mathrm{*}})\right)\\
 & =\lambda_{\textrm{nonzero}}\left(\left[\begin{array}{cc}
\varepsilon_{\mathrm{h},k}^{2}\boldsymbol{\Phi}_{k}^{b}\boldsymbol{\Phi}_{k}^{a}+\varepsilon_{\mathrm{g},k}^{2} & M\boldsymbol{\Phi}_{k}^{b}\end{array}\boldsymbol{\Phi}_{k}^{a}\right]\right)\\
 & =\lambda_{\textrm{nonzero}}\left(\left[\begin{array}{cc}
\varepsilon_{\mathrm{h},k}^{2}\boldsymbol{\Phi}_{k}^{a}\boldsymbol{\Phi}_{k}^{b}+\varepsilon_{\mathrm{g},k}^{2} & M\boldsymbol{\Phi}_{k}^{a}\boldsymbol{\Phi}_{k}^{b}\end{array}\right]\right)\\
 & =\lambda_{\textrm{nonzero}}\left(\left[\begin{array}{cc}
\varepsilon_{\mathrm{h},k}^{2}\boldsymbol{\Phi}_{k}+\varepsilon_{\mathrm{g},k}^{2} & M\boldsymbol{\Phi}_{k}\end{array}\right]\right)\\
 & =\lambda_{\textrm{nonzero}}\left(\boldsymbol{\Phi}_{k}\right)(\varepsilon_{\mathrm{h},k}^{2}+\varepsilon_{\mathrm{g},k}^{2}M).
\end{align*}
Then, we have
\begin{align*}
\widetilde{y}_{k}\mathbf{I}+\widetilde{\mathbf{U}}_{k}\succeq\mathbf{0} & \Longrightarrow\lambda(\widetilde{y}_{k}\mathbf{I})+\lambda(\widetilde{\mathbf{U}}_{k})\geq\mathbf{0}\\
 & \Longrightarrow\lambda(\widetilde{y}_{k}\mathbf{I})+(\varepsilon_{\mathrm{h},k}^{2}+\varepsilon_{\mathrm{g},k}^{2}M)\lambda(\boldsymbol{\Phi}_{k})\geq\mathbf{0}\\
 & \Longrightarrow\widetilde{y}_{k}\mathbf{I}+(\varepsilon_{\mathrm{h},k}^{2}+\varepsilon_{\mathrm{g},k}^{2}M)\boldsymbol{\Phi}_{k}\succeq\mathbf{0}
\end{align*}
using property: if $\lambda\left(\mathbf{A}\right)$ is the eigenvalue
of $\mathbf{A}$, then $\lambda\left(\mathbf{A}\right)+\sigma^{2}$
is the eigenvalue of $\mathbf{A}+\sigma^{2}\mathbf{I}$.

 \bibliographystyle{IEEEtran}
\bibliography{bibfile}

\begin{thebibliography}{10}
\providecommand{\url}[1]{#1}
\csname url@samestyle\endcsname
\providecommand{\newblock}{\relax}
\providecommand{\bibinfo}[2]{#2}
\providecommand{\BIBentrySTDinterwordspacing}{\spaceskip=0pt\relax}
\providecommand{\BIBentryALTinterwordstretchfactor}{4}
\providecommand{\BIBentryALTinterwordspacing}{\spaceskip=\fontdimen2\font plus
\BIBentryALTinterwordstretchfactor\fontdimen3\font minus
  \fontdimen4\font\relax}
\providecommand{\BIBforeignlanguage}[2]{{%
\expandafter\ifx\csname l@#1\endcsname\relax
\typeout{** WARNING: IEEEtran.bst: No hyphenation pattern has been}%
\typeout{** loaded for the language `#1'. Using the pattern for}%
\typeout{** the default language instead.}%
\else
\language=\csname l@#1\endcsname
\fi
#2}}
\providecommand{\BIBdecl}{\relax}
\BIBdecl

\bibitem{Marco-4}
E.~{Basar}, M.~{Di~Renzo}, J.~{De Rosny}, M.~{Debbah}, M.~{Alouini}, and
  R.~{Zhang}, ``Wireless communications through reconfigurable intelligent
  surfaces,'' \emph{IEEE Access}, vol.~7, pp. 116\,753--116\,773, 2019.

\bibitem{Marco-3}
M.~{Di Renzo}, K.~{Ntontin}, J.~{Song}, F.~H. {Danufane}, X.~{Qian},
  F.~{Lazarakis}, J.~{De Rosny}, D.~{Phan-Huy}, O.~{Simeone}, R.~{Zhang},
  M.~{Debbah}, G.~{Lerosey}, M.~{Fink}, S.~{Tretyakov}, and S.~{Shamai},
  ``{Reconfigurable intelligent surfaces vs. relaying: Differences,
  similarities, and performance comparison},'' \emph{IEEE Open J. Commun.
  Soc.}, vol.~1, pp. 798--807, 2020.

\bibitem{Xiaojun}
\BIBentryALTinterwordspacing
X.~Yuan, Y.-J. Zhang, Y.~Shi, W.~Yan, and H.~Liu,
  ``{Reconfigurable-intelligent-surface empowered 6G wireless communications:
  Challenges and opportunities},'' 2019. [Online]. Available:
  \url{https://arxiv.org/abs/2001.00364}
\BIBentrySTDinterwordspacing

\bibitem{Pan2019intelleget}
C.~Pan, H.~Ren, K.~Wang, M.~Elkashlan, A.~Nallanathan, J.~Wang, and L.~Hanzo,
  ``{Intelligent reflecting surface aided MIMO broadcasting for simultaneous
  wireless information and power transfer},'' \emph{IEEE J. Sel. Areas
  Commun.}, vol.~38, no.~8, pp. 1719--1734, Jun. 2020.

\bibitem{Pan2019multicell}
C.~Pan, H.~Ren, K.~Wang, W.~Xu, M.~Elkashlan, A.~Nallanathan, and L.~Hanzo,
  ``{Multicell MIMO communications relying on intelligent reflecting
  surface},'' \emph{IEEE Trans. Wireless Commun.}, vol.~19, no.~8, pp.
  5218--5233, May 2020.

\bibitem{Baitong2019}
T.~Bai, C.~Pan, Y.~Deng, M.~Elkashlan, and A.~Nallanathan, ``{Latency
  minimization for intelligent reflecting surface aided mobile edge
  computing},'' \emph{IEEE J. Sel. Areas Commun., early access}, 2020.

\bibitem{Marco-2}
\BIBentryALTinterwordspacing
H.~Han, J.~Zhao, D.~Niyato \emph{et~al.}, ``{Intelligent reflecting surface
  aided network: Power control for physical-layer broadcasting},'' 2019.
  [Online]. Available: \url{https://arxiv.org/abs/1910.14383}
\BIBentrySTDinterwordspacing

\bibitem{Gui2019IRS}
G.~Zhou, C.~Pan, H.~Ren, K.~Wang, and A.~Nallanathan, ``{Intelligent Reflecting
  Surface Aided Multigroup Multicast MISO Communication Systems},'' \emph{IEEE
  Trans. Signal Process.}, vol.~68, pp. 3236--3251, Apr. 2020.

\bibitem{Xianghao2009}
X.~{Yu}, D.~{Xu}, and R.~{Schober}, ``Enabling secure wireless communications
  via intelligent reflecting surfaces,'' \emph{2019 IEEE GLOBECOM}, pp. 1--6,
  Dec.

\bibitem{Shen2019secrecy}
H.~{Shen}, W.~{Xu}, S.~{Gong}, Z.~{He}, and C.~{Zhao}, ``{Secrecy rate
  maximization for intelligent reflecting surface assisted multi-antenna
  communications},'' \emph{IEEE Commun. Lett.}, vol.~23, no.~9, pp. 1488--1492,
  Jun 2019.

\bibitem{OFDM2019}
S.~Zhang and R.~Zhang, ``{Capacity characterization for intelligent reflecting
  surface aided MIMO communication},'' \emph{IEEE J. Sel. Areas Commun.},
  vol.~38, no.~8, pp. 1823--1838, Jun 2020.

\bibitem{taha2019enabling}
\BIBentryALTinterwordspacing
A.~Taha, M.~Alrabeiah, and A.~Alkhateeb, ``{Enabling large intelligent surfaces
  with compressive sensing and deep learning},'' 2019. [Online]. Available:
  \url{https://arxiv.org/abs/1904.10136}
\BIBentrySTDinterwordspacing

\bibitem{zhengyi-est}
\BIBentryALTinterwordspacing
Z.~Zhou, N.~Ge, Z.~Wang, and L.~Hanzo, ``{Joint transmit precoding and
  reconfigurable intelligent surface phase adjustment: A decomposition-aided
  channel estimation approach},'' 2019. [Online]. Available:
  \url{https://www.researchgate.net/publication/337824343}
\BIBentrySTDinterwordspacing

\bibitem{shuguang-IRS}
Z.~Wang, L.~Liu, and S.~Cui, ``{Channel estimation for intelligent reflecting
  surface assisted multiuser communications: Framework, algorithms, and
  analysis},'' \emph{IEEE Trans. Wireless Commun., early access}, 2020.

\bibitem{Peilan}
P.~Wang, J.~Fang, H.~Duan, and H.~Li, ``{Compressed channel estimation and
  joint beamforming for intelligent reflecting surface-assisted millimeter wave
  systems},'' \emph{IEEE Signal Processing Lett.}, vol.~27, pp. 905--909, May
  2020.

\bibitem{chenjie}
\BIBentryALTinterwordspacing
J.~Chen, Y.-C. Liang, H.~V. Cheng, and W.~Yu, ``{Channel estimation for
  reconfigurable intelligent surface aided multi-user MIMO systems},'' 2019.
  [Online]. Available: \url{https://arxiv.org/abs/1912.03619}
\BIBentrySTDinterwordspacing

\bibitem{Gui-letter}
G.~Zhou, C.~Pan, H.~Ren, K.~Wang, M.~{Di Renzo}, and A.~Nallanathan, ``{Robust
  beamforming design for intelligent reflecting surface aided MISO
  communication systems},'' \emph{IEEE Wireless Commun. Lett., early access},
  2020.

\bibitem{xianghao-robust}
X.~Yu, D.~Xu, Y.~Sun, D.~W.~K. Ng, and R.~Schober, ``{Robust and secure
  wireless communications via intelligent reflecting surfaces},'' \emph{IEEE J.
  Sel. Areas Commun., early access}, 2020.

\bibitem{Gui-globecom}
G.~Zhou, C.~Pan, H.~Ren, K.~Wang, and A.~Nallanathan, ``{Outage constrained
  transmission design for IRS-aided communications with imperfect cascaded
  channels},'' \emph{IEEE Globecom 2020, early access}.

\bibitem{bounded-channel}
M.~Botros and T.~N. Davidson, ``{Convex conic formulations of robust downlink
  precoder designs with quality of service constraints},'' \emph{IEEE J. Sel.
  Topics Signal Process.}, vol.~1, no.~4, pp. 714--724, Dec. 2007.

\bibitem{jun-channel}
J.~Zhang, M.~Kountouris, J.~G. Andrews, and R.~W. Heath, ``{Multimode
  transmission for the MIMO broadcast channel with imperfect channel state
  information},'' \emph{IEEE Trans. Commun.}, vol.~59, no.~3, pp. 803--814,
  Mar. 2011.

\bibitem{PCCP-boyd}
\BIBentryALTinterwordspacing
T.~Lipp and S.~Boyd, ``{Variations and extension of the convex-concave
  procedure},'' \emph{Optim. Eng.}, vol.~17, no.~2, pp. 263--287, 2016.
  [Online]. Available: \url{https://doi.org/10.1007/s11081-015-9294-x}
\BIBentrySTDinterwordspacing

\bibitem{boy-S-procedure}
S.~Boyd, L.~G. El, E.~Ferron, and V.~Balakrishnan, \emph{Linear matrix
  inequalities in system and control theory}.\hskip 1em plus 0.5em minus
  0.4em\relax Philadelphia, PA: SIAM, 1994.

\bibitem{Petersen-lemma}
I.~R. Petersen, ``{A stabilization algorithm for a class of uncertain linear
  systems},'' \emph{Syst. Contr. Lett.}, no.~8, pp. 351--357, 1987.

\bibitem{Gharavol2013TSP}
E.~A. {Gharavol} and E.~G. {Larsson}, ``{The sign-definiteness lemma and its
  applications to robust transceiver optimization for multiuser MIMO
  systems},'' \emph{IEEE Trans. Signal Process.}, vol.~61, no.~2, pp. 238--252,
  Jan. 2013.

\bibitem{book-convex}
S.~Boyd and L.~Vandenberghe, \emph{{Convex optimization}}.\hskip 1em plus 0.5em
  minus 0.4em\relax Cambridge Univ. Press, 2004.

\bibitem{CVX2018}
M.~Grant and S.~Boyd, ``{CVX: MATLAB software for disciplined convex
  programming},'' \emph{Version 2.1. [Online] http://cvxr.com/cvx, Dec. 2018.}

\bibitem{qingqing2019}
Q.~{Wu} and R.~{Zhang}, ``{Intelligent reflecting surface enhanced wireless
  network via joint active and passive beamforming},'' \emph{IEEE Trans.
  Wireless Commun.}, vol.~18, no.~11, pp. 5394--5409, Nov. 2019.

\bibitem{W-K-MA2014}
K.~{Wang}, A.~M. {So}, T.~{Chang}, W.~{Ma}, and C.~{Chi}, ``Outage constrained
  robust transmit optimization for multiuser miso downlinks: Tractable
  approximations by conic optimization,'' \emph{IEEE Trans. Signal Process.},
  vol.~62, no.~21, pp. 5690--5705, Nov. 2014.

\bibitem{Xinda2017}
X.-D. Zhang, \emph{{Matrix analysis and applications}}.\hskip 1em plus 0.5em
  minus 0.4em\relax Cambridge Univ. Press, 2017.

\bibitem{luo2010SDR}
Z.~{Luo}, W.~{Ma}, A.~M. {So}, Y.~{Ye}, and S.~{Zhang}, ``{Semidefinite
  relaxation of quadratic optimization problems},'' \emph{IEEE Signal Process.
  Mag.}, vol.~27, no.~3, pp. 20--34, May 2010.

\bibitem{Ben-Tal2001convex}
A.~Ben-Tal and A.~Nemirovski, \emph{{(Lectures on modern convex optimization:
  Analysis, algorithms, and engineering applications). Philadelphia, PA, USA:
  SIAM}}.\hskip 1em plus 0.5em minus 0.4em\relax MPSSIAM Ser. Optim., 2001.

\bibitem{3Gpp-channel}
3GPP, ``{Technical specification group radio access network; study on 3D
  channel model for LTE (release 12)},'' \emph{TR 36.873 V12.7.0}, Dec. 2017.

\end{thebibliography}

\end{document}